\documentclass{aa}  
\usepackage{graphicx}
\usepackage{textcomp}

\begin{document}

\title{S Coronae Australis -- a T Tauri Twin
          \thanks{Based on observations collected at the European Southern Observatory in Chile (75.C-0292) and the SMARTS facilities at the Cerro Tololo Inter-American Observatory Observatory in Chile. } }
         
\author{ G. F. Gahm\inst{1}
        \and P. P. Petrov\inst{2}
        \and L. V. Tambovtseva\inst{3}
        \and V. P. Grinin\inst{3,4}
        \and H. C. Stempels\inst{5}
         \and F. M. Walter\inst{6}   }
          
   \offprints{G. F. Gahm}

    \institute{Stockholm Observatory, AlbaNova University Centre, Stockholm University, SE-106 91 Stockholm, Sweden\\
              email: \mbox{gahm@astro.su.se}
        \and Crimean Astrophysical Observatory, Russian Academy of Sciences, 298409, Nauchny, Crimea
        \and Pulkovo Astronomical Observatory of the Russian Academy of Sciences, Pulkovskoe shosse 65, 196140 St. Petersburg, Russia
         \and Sobolev Astronomical Institute, St. Petersburg State University, St. Petersburg, Russia
        \and Department of Physics and Astronomy, Uppsala University, Box 516, SE-75120 Uppsala, Sweden  
        \and  Department of Physics \& Astronomy, Stony Brook University, Stony Brook NY 11794 -3800, USA }
        
   \date{}

   

 \abstract
   {The star S CrA is a tight visual binary consisting of two classical T Tauri stars. Both components are outstanding regarding their spectral characteristics and brightness variations.} 
   {Our aim is to explore the extraordinary spectral features seen in these stars, derive stellar parameters, define spectral signatures of accreting gas and winds, estimate the inclinations of the disks, and to match numerical models with observed properties.  }
   {High-resolution spectra were collected of each component over several nights at the European Southern Observatory (ESO) combined with photometric observations covering several years in UBVRI with the SMARTS telescope. The models developed include magnetospheric accretion and a disk wind.}
   {Both stars undergo large variation in brightness, $\geq$ 2 mag in V band. The variations are caused mainly by variable foreground extinction from small-size dust grains, which may be carried along with the accreting gas. The photospheric absorption lines are washed out by superimposed continuous and line emission, and this veiling becomes occasionally exceptionally high. Nevertheless, we extracted the stellar spectra and found that both stars are very similar with regard to stellar parameters ($T_{eff}$, log $g$, $v$ sin $i$, mass, radius, luminosity). The rotational periods, inferred from velocity shifts in lines originating in surface areas off-set from the pole, are also similar. Combined with the $v$ sin $i$:s related inclinations were obtained, which agree well with those derived from our model simulations of Balmer line profiles: $\sim$ 65$\degr$ for both stars. At this orientation the trajectories of infalling gas just above the stellar surfaces are parallel to the line-of-sight, and accordingly we observe extended red-shifted absorption components extending to + 380 km s$^{-1}$, the estimated free-fall velocity at the surface. Rates of accretion and mass loss were obtained from the models.  }
  {The two stars are remarkably similar, and S CrA can be regarded as a T Tauri twin. The components differ, however, in terms of degree of veiling and emission line profiles. We have found a good match between observed signatures of accreting gas, wind features, and rotational velocities with those resulting from our modelling for inclinations of $\sim$ 65$\degr$. These inclinations differ from those derived from interferometric near-infrared (NIR) spectroscopy, and possible causes for this puzzling discrepancy are discussed.}

  \keywords{Stars: variables: T Tauri -- stellar parameters, accretion, winds -- stars: individual: S CrA}
  
 \maketitle
%

\section{Introduction}
\label{sec:intro}

Classical T Tauri stars (cTTS) are pre-main sequence stars (PMSs) of low mass characterized by their rich emission line spectra with 
signatures of both accreting and outflowing gas. Their complex spectra are understood as a result of magnetospheric accretion of material from circumstellar disks in combination with emission from stellar winds and also winds arising in the surrounding disk. Spectral types can be deduced from the photospheric lines, but in cTTSs the stellar absorption lines are weakened due to overlying excess emission, an effect known as veiling, which has been related to continuous emission from the shocked region below the accretion streams close to the stellar surface. For reviews of properties and models of cTTS, see Petrov (\cite{pet03}), Bouvier et al. (\cite{bou07}), and Hartmann et al. (\cite{har16}).

Some cTTSs have exceptionally strong and broad absorption components flanking the red wings of certain emission lines, called inverse P Cygni profiles. Such stars are sometimes referred to as YY Orionis stars after Walker's (\cite{wal72}) spectroscopic survey of objects in two young clusters. The red-shifted components originate in gas falling towards the star with velocities of  $>$ 100 km s$^{-1}$.

One PMS system with outstanding YY Ori characteristics is the 1.$\arcsec3$ visual binary S Coronae Australis. At a distance of 130 pc (Reipurth \& Zinnecker \cite{rei93}), this corresponds to a projected separation of 169 AU if physical. Both components, usually designated as NW (north-west) and SE (south-east), have very rich emission line spectra and vary in brightness by up to two magnitudes in the photometric $V$ band on timescales of hours and days and become redder with decreasing brightness (Walter \& Miner 2005). 

A number of spectroscopic studies of the YY Ori star S~CrA followed the survey by Walker (see  Appenzeller et al. \cite{app86}, and references therein; Krautter et al. \cite{kra90}; Graham \cite{gra92}; Hamann \& Persson \cite{ham92}) demonstrating that the red-shifted absorption components are present in different types of lines, and in cases when repeated observations were made, they were found to vary in strength. However, the NW and SE components were not clearly resolved until Prato et al. (\cite{pra03}), Takami et al. (\cite{tak03}), and Carmona et al. (\cite{car07}) presented individual spectra.

Over the last decade, more detailed analyses of the spectral properties of both components have been made. On the basis of high-resolution spectra obtained of each component at the Very Large Telescope (VLT at the European Southern Observatory (ESO)), Gahm et al. (\cite{gah08}) focussed on the nature of the veiling, which is variable and on occasion extraordinarily strong. It was demonstrated that when the veiling becomes high, the photospheric lines are filled in by narrow emission lines unrelated to any continuous emission from shocked regions. This phenomenon was related to accretion-powered chromospheric emission in Petrov et al. (\cite{pet11}), and a similar concept was adopted in the models by Dodin \& Lamzin (\cite{dod12}) and Dodin et al. (\cite{dod13}). 

Emission lines with prominent red-shifted absorption were further studied in Petrov et al. (\cite{pet14}), hereafter called Paper I) who found that the radial velocity of the deepest part depends on the excitation and ionization state of the line. On occasion, these components were exceptionally broad and strong in S CrA SE and it was demonstrated that they could be matched with an accretion flow as depicted in the standard magnetospheric models assuming a dipole magnetic field structure. Gas at different density and temperature is found at different distances from the stellar surface, resulting in the observed velocity pattern when viewed along a sight line through the magnetosphere. 

The circumstellar disks and associated spectral signatures and disk winds have been discussed in Kessler-Silacci et al. (\cite{kes06}), Schegerer et al. (\cite{sch09}), Pontoppidan et al. (\cite{pon11}), Bast et al. (\cite{bas11}), Vural et al. (\cite{vur12}), and Meeus et al. (\cite{mee13}). Recently Garcia Lopez et al. ({\cite{gar17}, GRAVITY collaboration) resolved the inner gaseous disk in both components from interferometric observations of their near-infrared (NIR) spectra. They found signatures of both accretion flows and jet streams in the Br$\gamma$ and $\ion{He}{i}$ lines and obtained disk inclinations that are rather similar in the two stars, 28 and 22 degrees, implying that {\bf the systems are seen} nearly face-on. Finally, the Herbig Haro objects HH 82 and 729 appear to be related to S CrA (Reipurth \& Graham \cite{rei88}; Wang et al. \cite{wan04}).   

In the present paper we revisit the material of high-resolution spectra obtained at VLT of both S CrA NW and SE with the purpose of deriving stellar parameters for the two stars and to disentangle the different spectral signatures of in-flowing and out-flowing matter in an attempt to match these with numerical models of the magnetospheric structures envisioned for the two stars. 

The observations and reductions are described in Section~\ref{sec:obs}. In Section~\ref{sec:results} we spot-light different spectral features that are  related to stellar winds and accretion streams, and which form the basis for our theoretical models of the objects presented in Section~\ref{sec:models}. The results are further discussed in Section~\ref{sec:discussion}, and we end with a summary in Section~\ref{sec:conclusions}.

\section{Observations}
\label{sec:obs}

Eleven high-resolution spectra of S CrA were obtained between August 14 and 17, 2005 (MJD 24553597.03 to 24553600.02), with the UVES spectrograph in the dichroic-mode on the 8 m VLT/UT2 at ESO, Chile. The stars were aligned in the slit and were perfectly spatially resolved, therefore the two components were extracted individually.  The spectra have a high signal-to-noise ratio, S/N $\ga 100$, a spectral resolution of $R \approx 60\,000$, and a wavelength coverage from 3500\,to\,6700 {\AA}. For more information on data reductions see Stempels et al. (\cite{ste07}). In addition, we have compiled a long-term record of the two S CrA components with $UBVRI/JHK$ photometry from 2003\,to\,2007, using the ANDICAM dual-channel imager on the SMARTS 1.3 m telescope at Cerro Tololo.

\section{Results}
\label{sec:results}

\subsection{Photometry}
\label{photometry}

Both components in S CrA  (referred to as NW and SE below) are strongly variable and become redder when fainter as is evident from our long-term photometry with SMARTS presented in the upper panel of Fig.~\ref{photometry}. Most of the time, NW is brighter than SE, on the average by 0.$^m$6 in the V band. The lower panel of Fig.~\ref{photometry} shows how the two stars varied in the I and B bands over the period when the VLT spectra were collected. The variations were moderate in the I band, but in B, changes of almost a magnitude occurred on timescales of one day in both stars.

The veiling changed dramatically over this period, especially in SE, but these changes are not tightly related to the brightness changes, because a contribution of varying line emission dominates when the veiling is high (Gahm et al. {\cite{gah08}). Instead, the main cause of variability is obscuration by passing circumstellar dust clouds in the line-of-sight. The slopes of the observed reddening of the stars in  Fig.~\ref{photometry} is consistent with a normal reddening law, so the dust grains must be small.   

Both stars show extraordinarily large variations in the optical bands. There is no clear periodicity in these changes judging from our long-term photometric monitoring, and large changes can occur over days as well as over weeks. Variations due to extinction events are not unusual among the TTSs (e.g., Herbst et al.  \cite{her94}; Guo et al. \cite{guo18}, and references therein).   

\begin{figure}
\centerline{\resizebox{7.6cm}{!}{\includegraphics{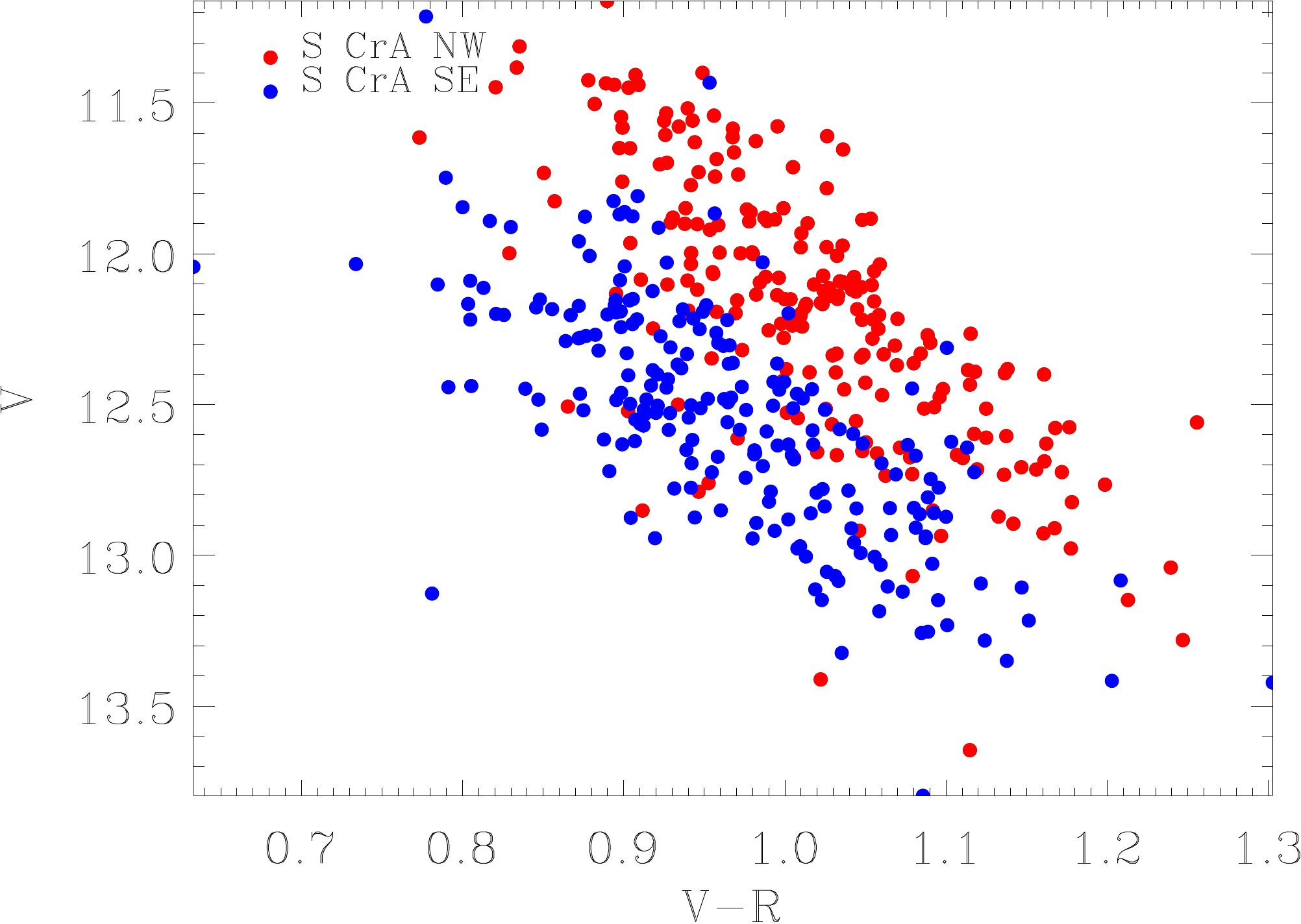}}}
\centerline{\resizebox{7.1cm}{!}{\includegraphics{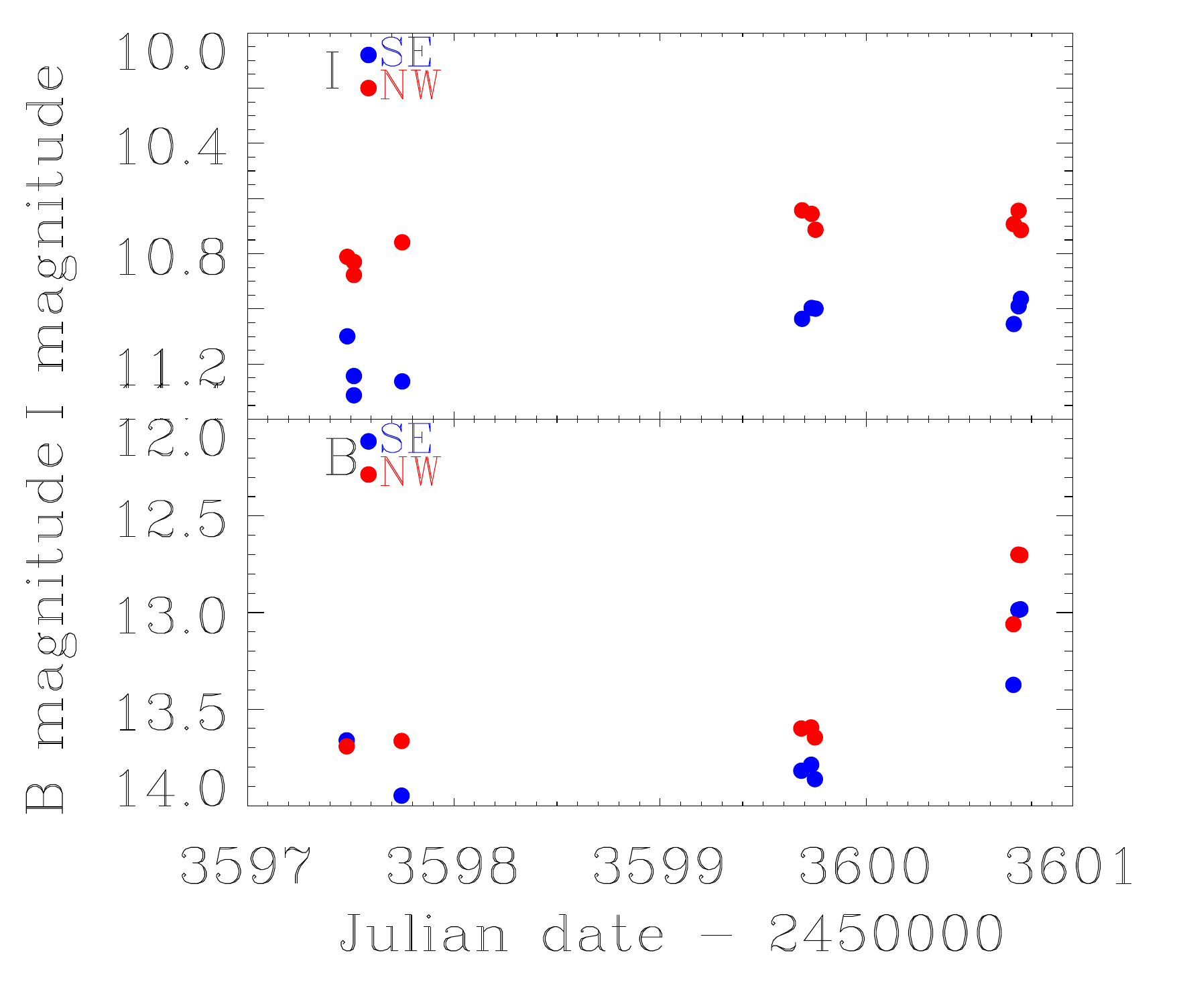}}}
\caption{Upper panel: Colour-magnitude diagram from the long-term monitoring of S CrA (blue: S CrA SE, red: S CrA NW)  Lower panel: Individual I and B magnitudes observed over the period of spectral observations.}
\label{photometry}
\end{figure}

\subsection{Stellar lines and parameters}
\label{sec:photospheric}

\begin{figure}
\centerline{\resizebox{9cm}{!}{\includegraphics{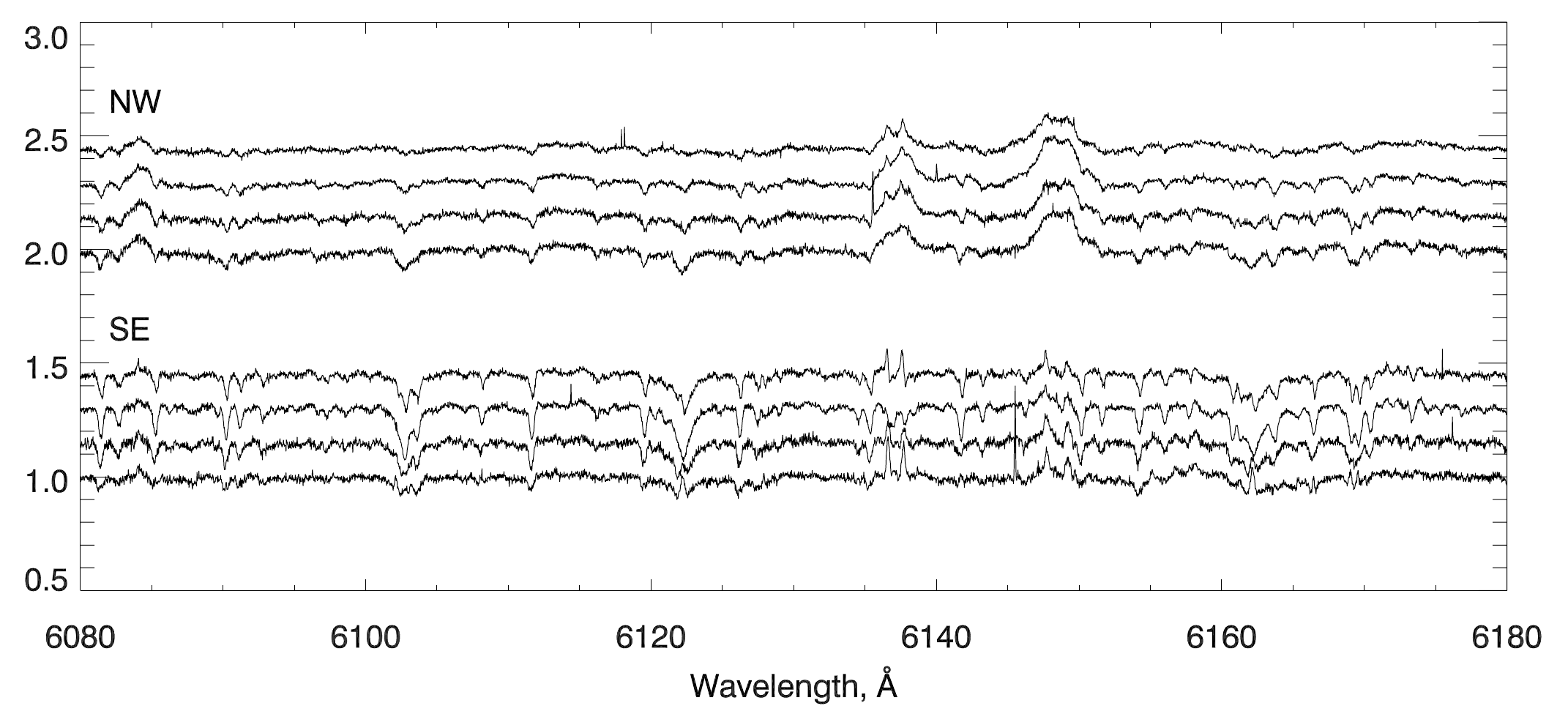}}}
\centerline{\resizebox{9cm}{!}{\includegraphics{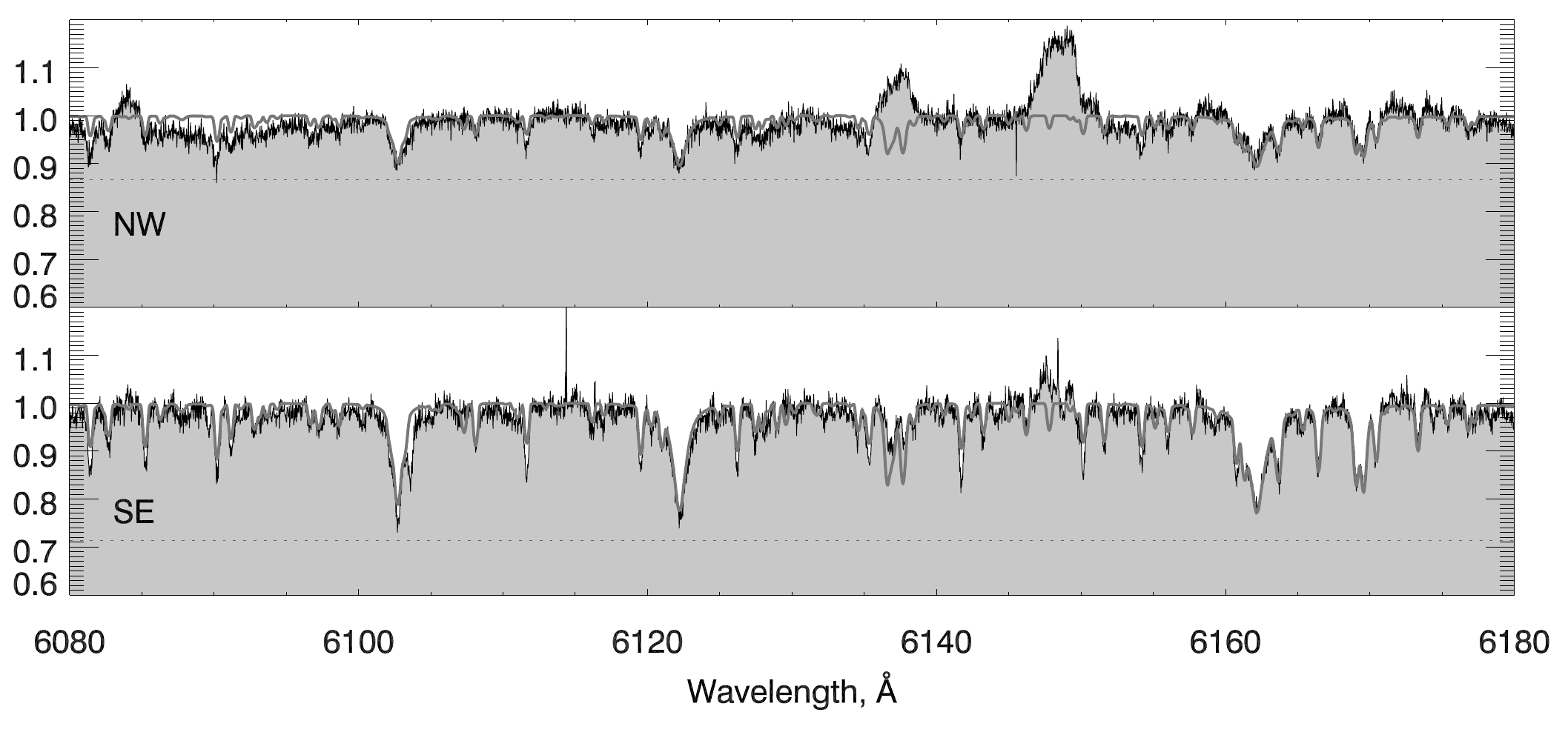}}}
\caption{Upper panel: Spectra of the stars obtained during the 4 nights over a spectral region relatively free from emission lines and veiling used to derive stellar parameters (time running upwards). Lower panel: Showing the good agreements between observed (black) and synthetic (grey) spectra. Different contributions from a continuous source of veiling is added (horizontal dashed lines). }
\label{photosphere}
\end{figure}  

The photospheric lines are completely obliterated by superimposed emission in the blue spectral region in both components. The top panel in Fig.~\ref{photosphere} shows spectra of NW and SE obtained on the four consecutive nights over the region 6000\,--\,6180 \AA, one of the few spectral regions that is not contaminated with line emission. We determined the fundamental stellar parameters of both components by calculating synthetic spectra with the SME software (Valenti \& Piskunov \cite{val96}; Piskunov \& Valenti \cite{pis17}) extended with routines to compensate for the level of veiling. The radiative transfer calculations in SME were performed using the Kurucz (\cite{kur93}) model atmospheres and atomic data from the VALD database (Piskunov et al. \cite{pis95}; Kupka et al. \cite{kup99}). We found that the photospheric spectra of the two stars are extremely similar, with identical $v$ sin $i$, $T_{eff}$, and log~$g$. 

The photospheric lines are weaker due to veiling compared to normal stars of the same spectral type. The veiling factor (VF) is defined as the ratio of an assumed excess continuum to the stellar continuum.  The lower panel in Fig.~\ref{photosphere} shows the excellent agreement between the observed (black line) and synthetic (grey line) spectra. The only difference between the two components is the added level of veiling, here depicted as a continuous excess emission (horizontal dashed lines). However, as described in Section~\ref{sec:intro}, narrow emission lines entering inside the photospheric lines also contribute to the veiling. This effect of line-dependent veiling in highly accreting stars was studied in more detail by Rei et al. (\cite{re18}).

In Table~\ref{parameters} we summarize the so-derived stellar parameters, which turn out to be remarkably similar for the two components. From the evolutionary tracks by Siess et al. (\cite{sie00}), assuming $T_{eff}$ = 4250~K and $L_*$ = 0.7~L$_{\odot}$ for each component, we find $M_*$ = 1~M$_{\odot}$ and $R_*$ = 1.4~R$_{\odot}$ for both stars. This is consistent with the derived value of log g = 4.0. {\it Hence, S Cr A is an example of a stellar twin}.  However, as can be seen in the table, the mean veiling is much higher in NW than in SE.

\begin{table}
\caption{Stellar parameters.}
\label{parameters}
\begin{tabular}{lcc} 
\hline
     \noalign{\smallskip}
  &     SE component & NW component\\
\hline
     \noalign{\smallskip}
$v_{rad}$ (km s$^{-1}$) & - 0.5 $\pm$ 3.9  & + 0.9 $\pm$ 2.5 \\
$v$ sin $i$ (km s$^{-1}$) &  12 & 12 \\
$T_{eff}$ (K) & 4250 & 4250 \\
Spectral type & K6 IV & K6 IV \\
log $g$ & 4.0  & 4.0 \\
mean veiling factor & 3.3 & 8.3 \\
     \noalign{\smallskip}
\hline
\end{tabular}
\end{table}

\subsection{Spectral signatures of gas flows}
\label{subsec:flows}

\begin{figure}
\centerline{\resizebox{9.3cm}{!}{\includegraphics{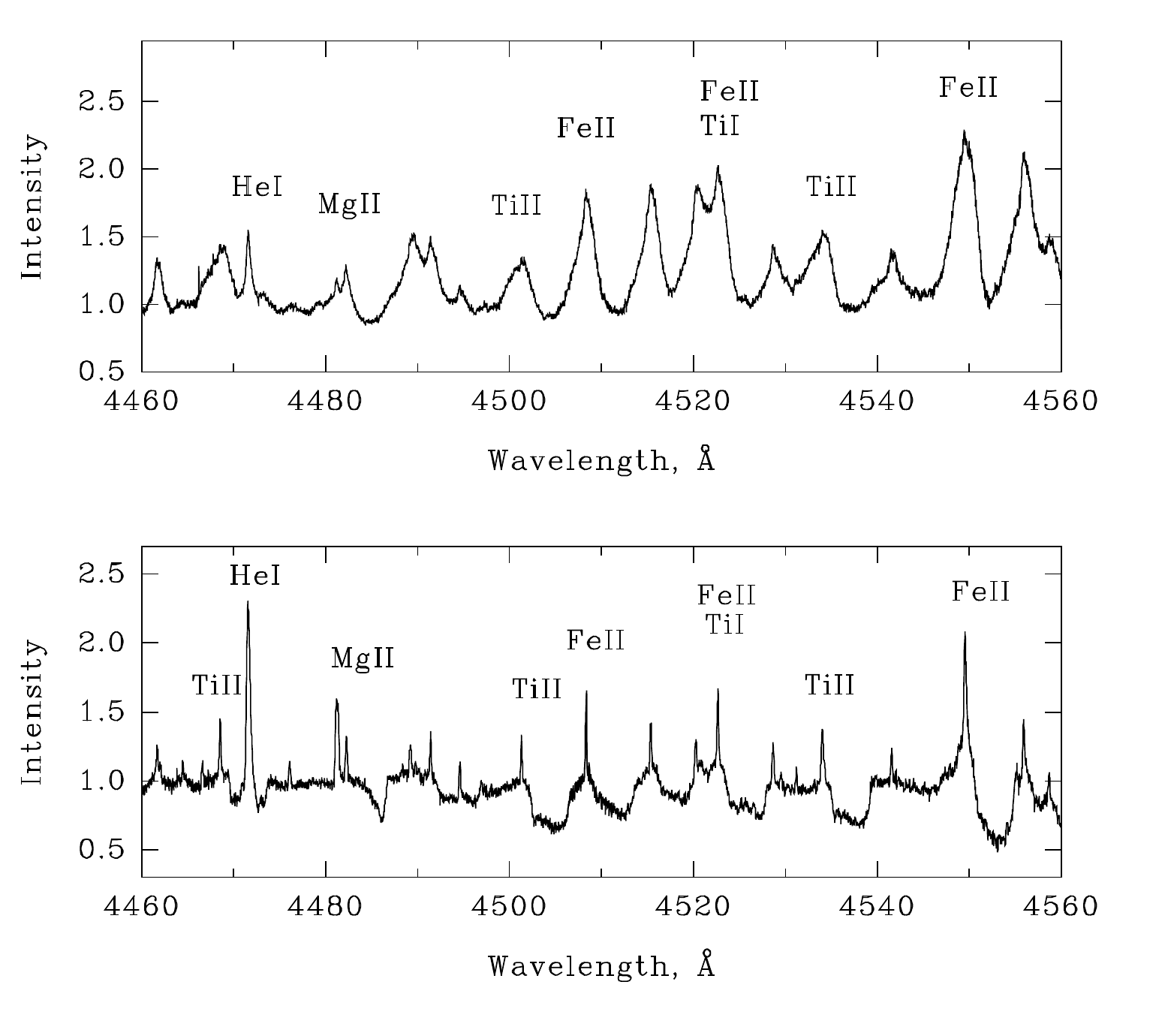}}}
\caption{Segments of spectra of S CrA NW (upper) with broad emission lines and S~CrA~SE (lower) with strong red-shifted absorption components. Narrow emission lines of metals are present in both stars. }
\label{4460}
\end{figure} 

\begin{figure}
\centerline{\resizebox{9cm}{!}{\includegraphics{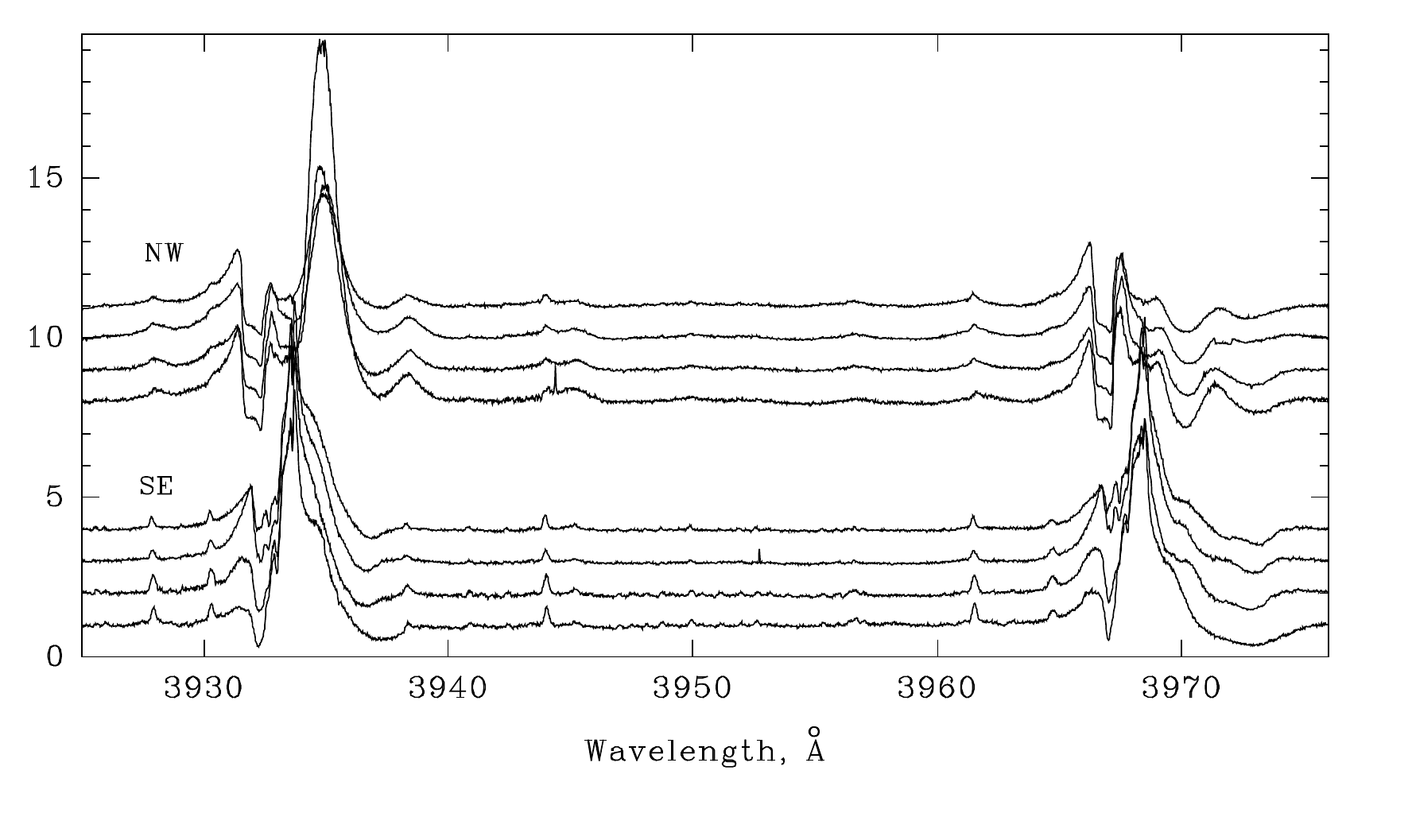}}}
\includegraphics[angle=00, width=9cm]{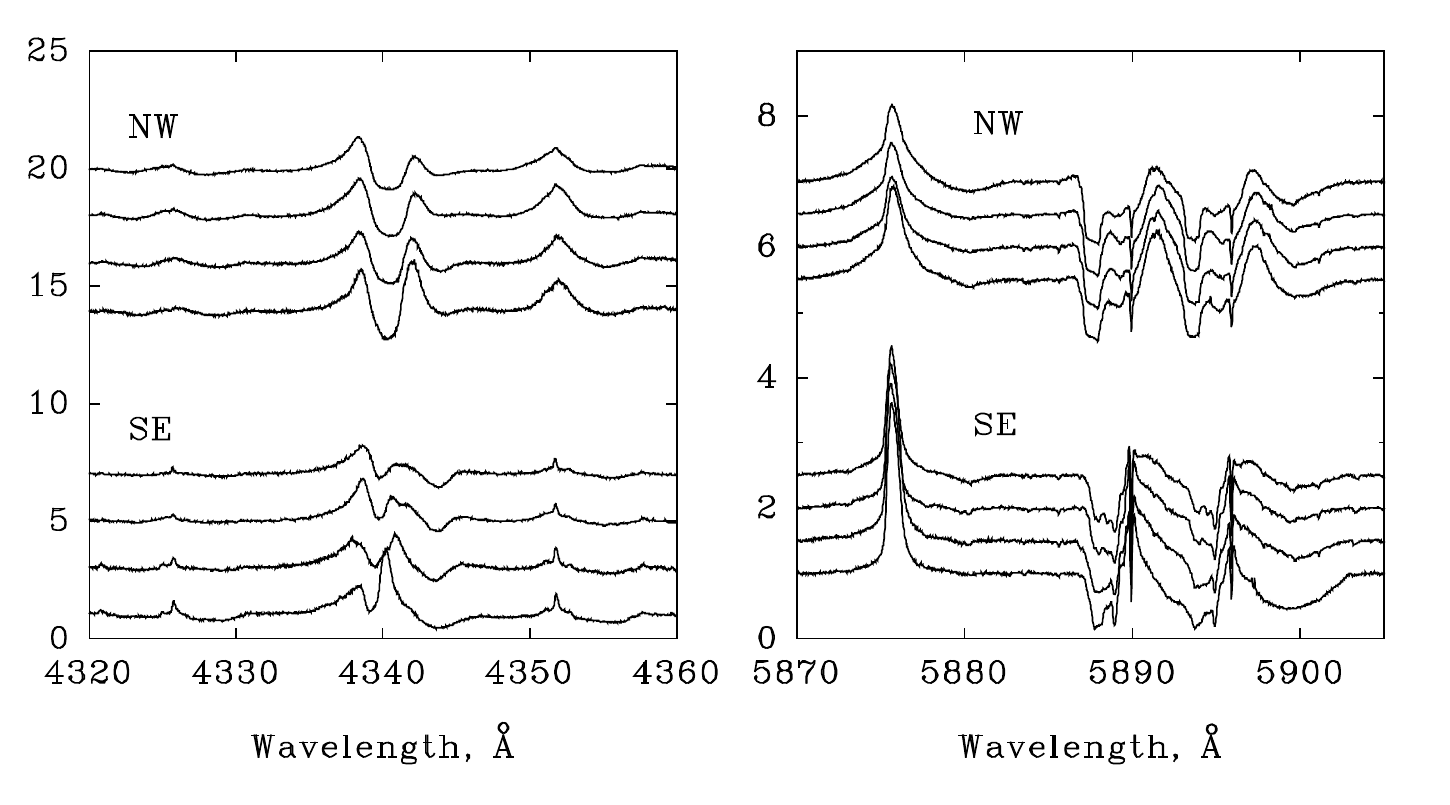}
\caption{Spectral segments of the two stars taken during 4 nights over the \ion{Ca}{ii} H \& K (upper panel) and H$\gamma$ and $\ion{He}{i}$ plus $\ion{Na}{i}$ D lines (lower panel). Intensities are normalized to the continuum level and time is running upwards. }
\label{IPC}
\end{figure}  

Emission lines of H and, $\ion{He}{i}$, and neutral and ionized metals are prominent in the spectra of both stars. Figure~\ref{4460} shows segments of spectra from August 14. The metal lines are broad in NW ($ \sim$ 100 km s$^{-1}$) on top of which narrow emission lines are present in several lines. In SE, the metal lines are as narrow as the photospheric lines and show velocity changes in anti-phase with the photospheric lines as reported in Petrov et al. (\cite{pet11}) suggesting that the line emission is confined to a certain area on the stellar surface, which is not aligned with the axis of rotation. Strong red-shifted absorption components flank the metal lines in SE.

Spectra taken over the four nights covering the \ion{Ca}{ii} H \& K, H$\gamma$, $\ion{He}{i}$ D3, and $\ion{Na}{i}$ D lines are shown in Fig.~\ref{IPC}. Red- and blue-shifted absorption components are prominent in both stars. The red-shifted components were very stable in NW during the period, while in SE they were highly variable and on occasions extremely strong. The emission components vary on time scales of a day. We note that in the $\ion{Ca}{ii}$ H profile of NW the red peak is depressed by the central absorption of H$\epsilon$, and the blue peak of $\ion{Ca}{ii}$ H is also depressed by absorption in the cool wind. As a result, the H line becomes considerably weaker than the K line. 

\begin{figure}
\centering
\includegraphics[angle=00, width=8.5cm]{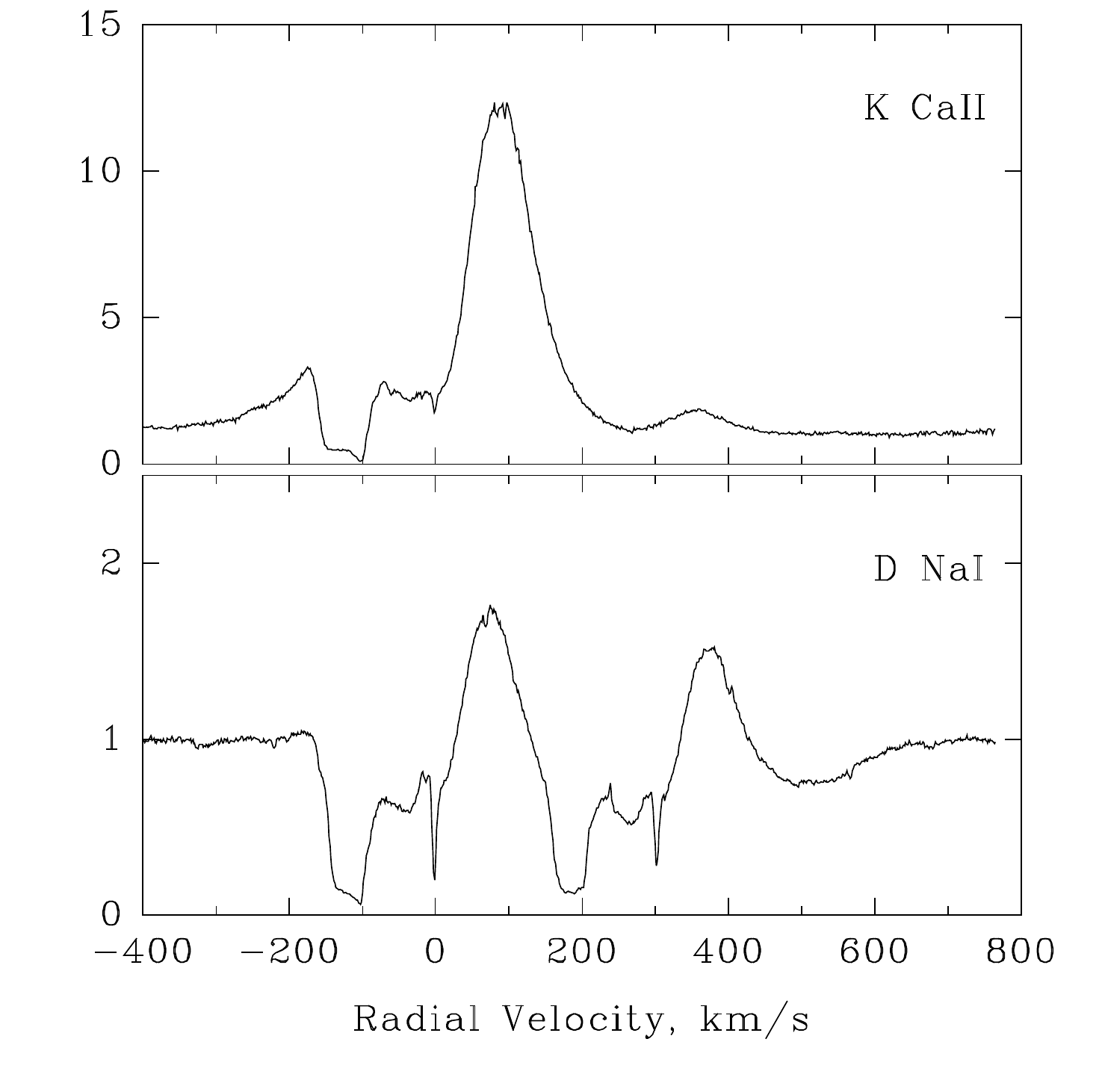}
\caption{The \ion{Ca}{ii} K {\bf and} $\ion{Na}{i}$ D lines in S CrA NW show strong blue-shifted absorption components in addition to narrow interstellar lines (spectra from August 14). Velocities are astro-centric in this and similar figures.}
\label{na/ca}
\end{figure} 

The presence of both blue- and red-shifted absorption components in the Balmer lines were recognised early by, for example,  Rydgren (\cite{ryd77}) and Edwards (\cite{edw79}) and at the time it was discussed whether the blue-shifted components could arise in winds as well as in accretion flows. The blue-shifted components in the H \& K and D lines reach below the continuum in both stars and must be related to a wind. Both the strength and velocity shifts of these absorption features are larger in NW, indicating a higher mass-loss rate in this star. In NW, neutral gas moves at 120 km s$^{-1}$ relative to the star as seen in Fig.~\ref{na/ca}. Obviously, this absorption line originates from gas that has been launched at high speed into the line-of-sight. In addition, narrow interstellar absorption components are present.

Figure~\ref{balmer} shows the observed Balmer line profiles in NW from spectra taken on August 14 and in SE on August 16, the dates of minimum veiling for each star. Common for both stars are the double-peaked profiles with a broad `central' absorption component, strongest in H$\alpha$, and a broad red-shifted absorption component in the higher Balmer lines. The strong central absorption component in NW is typical for a disk wind observed at high inclinations of the disk relative to the sky plane (e.g. Kurosawa et al.~\cite{kur11}). 

In the case of NW, the deepest part of the wind feature is at about - 70 km\,s$^{-1}$ in H$\alpha$, - 20 km\,s$^{-1}$ in H$\beta$, and is close to zero velocity in H$\gamma$ and H$\delta$. This  reflects the outward accelerated wind since in the higher Balmer lines the absorption saturates in the inner and densest part of the wind, while in H$\alpha$ the absorption is saturated already in regions of  lower density and at larger expansion velocities. 

\begin{figure}
\centering
\includegraphics[angle=00, width=4cm]{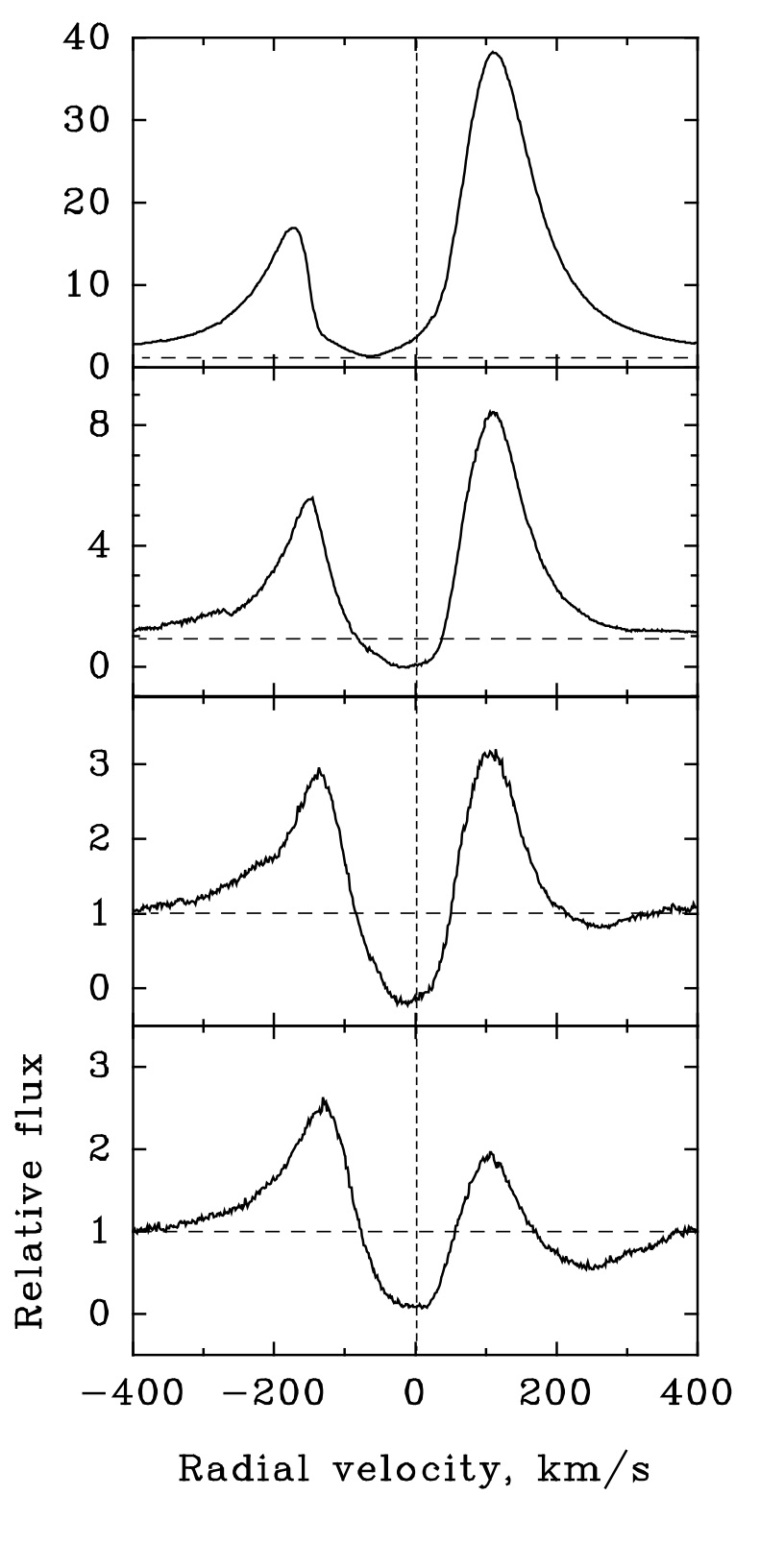}
\includegraphics[angle=00, width=4.5cm]{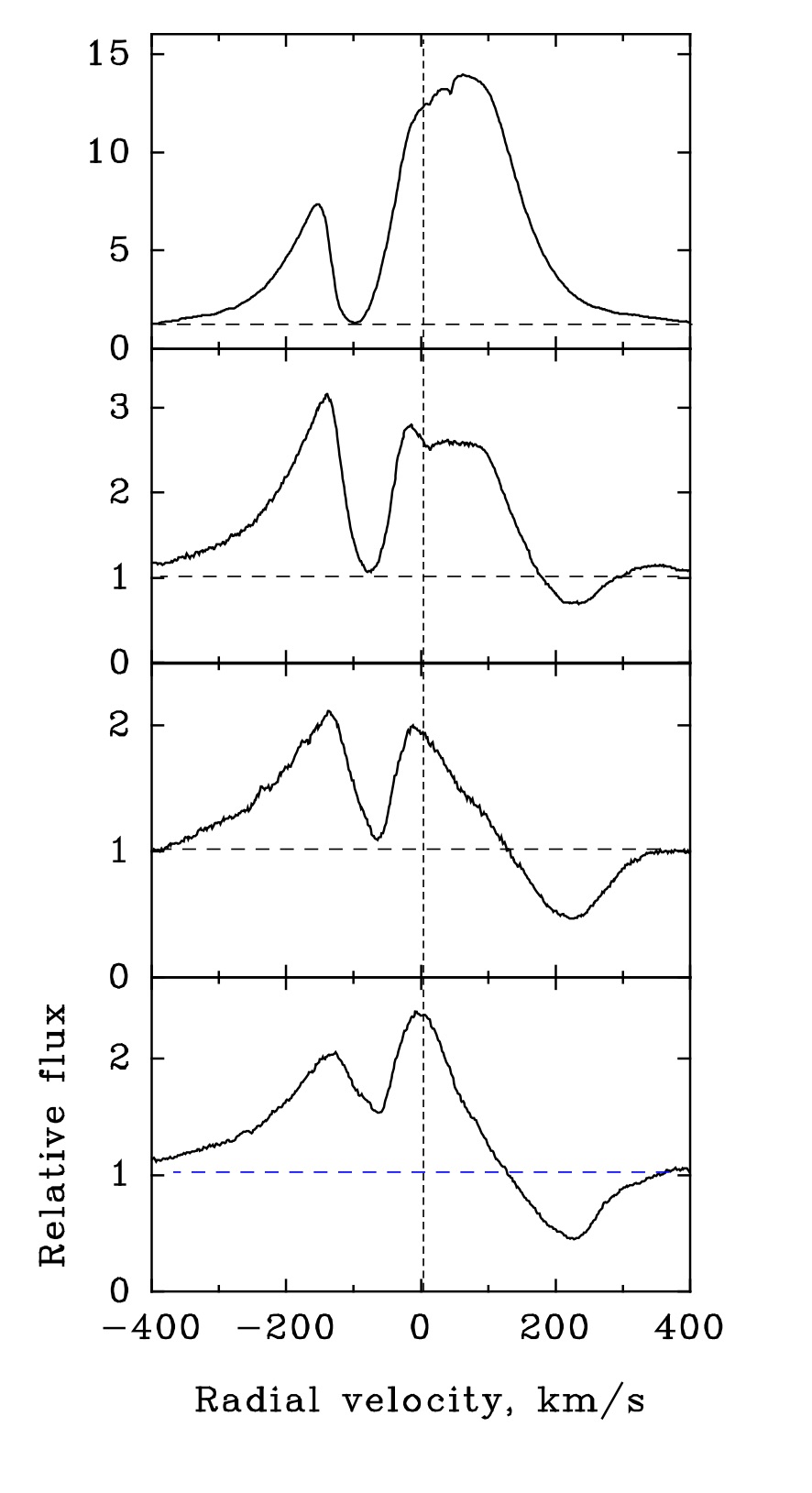}
\caption{Balmer line profiles. Left: S CrA NW (August 14). Right: S~CrA~SE (August 16). From top to bottom H$\alpha$, H$\beta$, H$\gamma$, and H$\delta$ with the continuum level marked with dashed lines. }
\label{balmer}
\end{figure} 

The emission lines of He and metals are composed of one broad and one narrow  component. According to Kwan \& Fischer~(\cite{kwa11}) and Beristain et al.~(\cite{ber98}) the broad component can originate in the magnetosphere loaded with accreting gas characterized by $T_e$ = 7000 -- 8000~K and log\,$N_e$ = 10 -- 12, whilst the narrow components are usually attributed to the post-shocked gas with physical conditions close to those in the solar chromosphere, $T_e$ $\sim$ 10$^4$, log\,$N_e$ $\approx$ 12. The stellar wind is characterized by high temperature and low density, $T_e$~$\ge$~10$^{4}$~K, log\,$N_e$~=~9\,--\,11.

Regarding the red-shifted absorption components, several examples of line profiles observed in SE were shown in Paper I. In the August 14 spectrum of SE, the veiling is very high, and on this occasion many lines of different elements at different ionization stages show prominent inverse P Cyg profiles. The physical conditions along the accretion funnels could then be derived from a simple dipole model, and mass-loss rates of the order of 10$^{-7}$ M$_\odot$/yr were found. This result agreed with expected rates derived from the equivalent widths of Hydrogen and Helium emission lines, using the  empirical relations between line luminosity and mass accretion rate (e.g. Fang et al.~\cite{fan09}). From these relations, one would conclude that the mass-loss rate is twice as large in NW compared to SE. However, the accretion components are stronger in SE for possible reasons discussed in Section~\ref{sec:discussion}.

\section{Model calculations}
\label{sec:models}

In order to determine the main physical processes in the nearest environment of both components in S CrA and to estimate inclination angles, we have modelled the radiation coming from different regions in the environments of the stars, including one region of magnetospheric accretion (MA) and one with a disk wind (DW). First, we calculated the hydrogen lines of the Balmer series for each region separately, and then we combined these two regions, and calculated line profiles within the hybrid model as depicted in Fig.~\ref{fig:sketch}. We took into account all foreground absorption of the radiation from the star and the accretion region by the disk wind. Below, we briefly describe the main assumptions of both models regarding kinematics and temperature regimes. For a full description of these models, see Grinin \& Tambovtseva (\cite{gri11}) and Tambovtseva et al. (\cite{tam16}).

\begin{figure}
\centering
\includegraphics[width=8cm,angle=-00]{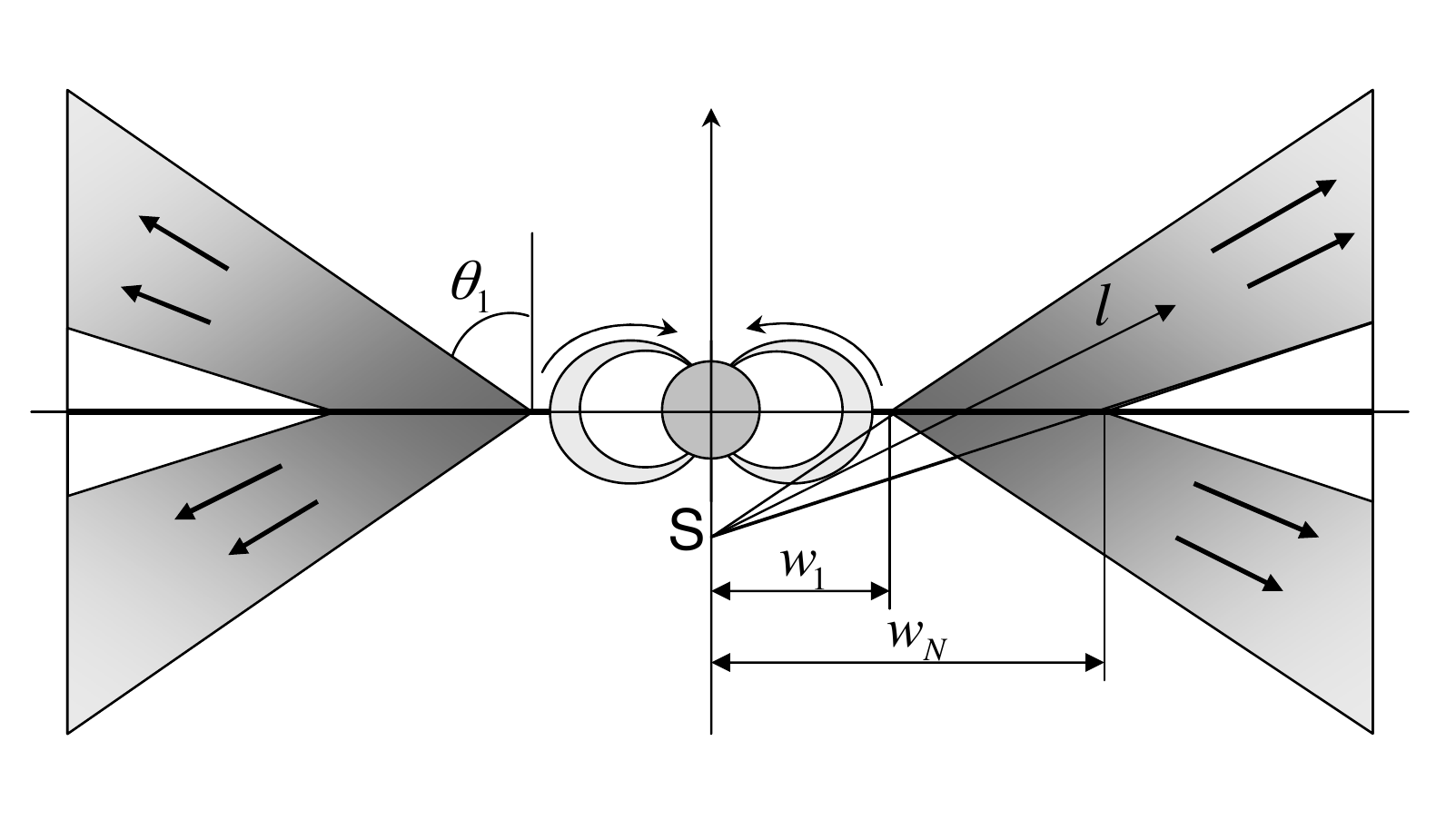}
\caption{Sketch of the hybrid model geometry assumed in the
present calculations (not to scale). The inclination angle is measured from the
symmetry axis (S). }
\label{fig:sketch}
\end{figure}

\subsection{Disk wind}
\label{subsec:disk}

The disk wind model is based on the theory of a magneto-centrifugal disk wind developed for the accretion disks surrounding black holes by Blandford \& Payne (\cite{bla82}) and was used to study the accretion disks near young stellar objects by Pudritz and Norman (\cite{pud86}). Magneto-centrifugal disk winds in T Tauri stars have been investigated in many works (e.g. K\"{o}nigl \& Pudritz  \cite{koe00}; Ferreira \cite{fer13}, and references therein). Safier (\cite{saf93}) and Garcia et al. (\cite{gar01}) argued that the rising gas is rapidly heated by ambipolar diffusion up to $\sim$ 10 000 K. Observations of jets from TTSs favour models of extended centrifugally driven winds (Ferreira et al. \cite{fer06}).

The geometry and kinematics of the disk wind obtained from solutions of the MHD equations can be reconstructed with the method of parametrization (e.g. Kurosawa et al. \cite{kur06}). In short, the main properties of the disk wind model are as follows. We divide the disk wind region by a number of streamlines extending from one point. For simplicity we assume that the disk wind consists of hydrogen atoms with a constant temperature 
(10 000 K) along the streamlines. Since the gas temperature just above the disk surface is not high enough to excite the hydrogen line
emission, we exclude the base of the disk wind in the calculation and use a scaling factor for the temperature along each streamline.

The main model parameters are as follows. The footpoints of the disk, $w_1$ and $w_N$, refer to the wind launch region for the first and last streamlines, respectively. $\theta_1$ is the half opening angle between the first streamline and the vertical axis (Fig.~\ref{fig:sketch}). The poloidal velocity component $v(l)$ changes along the streamlines, as given by

\begin{equation}\label{v}
    v(l) = v_0 + (v_{\infty} - v_0)\,(1 - l_i/l)^{\beta}\,.
\end{equation}
Here, $v_0$ and $v_{\infty}$ are the initial and terminal velocities and $\beta$ is a parameter. We assume that $v_0$ is the sound velocity in the disk wind. $v_\infty$\,=$fu_K(w_i)$, where $u_K(w_i)$ is the Keplerian velocity at the footpoint $w_i$ on the disk surface for the $i$th streamline, and $f$ is a scale factor.

The tangential velocity $u(w)$ changes with a cylindric radius, $w$, as follows (Grinin et al. \cite{gri17}): 

In the corotation zone ($u(w)/u_K(w_i)) \leq f_c$)  

\begin{equation}\label{u1}
    u(w)=u_K(w_i)(w/w_i).
\end{equation}

In the zone of conservation of angular momentum
\begin{equation}\label{u2}
    u(w)=u_K(w_i) f_c/(w/w_i),
\end{equation}
where $f_c$ is a parameter.

Two additional parameters enter, namely the mass-loss rate $\dot M_{w}$ and $\gamma$. The latter regulates the mass loading among the streamlines.
We assumed that the accretion disk is transparent to radiation up to the distance of a sublimation radius. Beyond the sublimation radius, the disk is opaque obscuring the disk wind moving away from the observer.

\subsection{Magnetospheric accretion model}
\label{subsec:magnetospheric}

We consider models in which the accreting matter falls onto the stellar surface radially within two cones. The cones extend to a radial distance $R_{out}$. We neglect the outer regions of MA, which contribute much less to the hydrogen line emission than the inner parts because of the lower gas temperatures. The same scheme of accretion was used by Calvet and Hartmann (\cite{cal92}).

The underlying assumptions are in short: the accreting gas flows from the outer radius $R_{out}$ to the stellar surface within two cones with half-opening angles $\theta_{in}$ and $\theta_{out}$ and described by one radial ($v$) and one tangential ($u$) velocity component.

\begin{equation}\label{vms}
    v(r) = \sqrt{v_*^2 + v_{esc}^2 \left(\frac{R_*}{r}- \frac {R_*}{R_{out}}\right)},
\end{equation}

\begin{equation}\label{ums}
    u(w) = U_* \frac{w}{R_*},
\end{equation}
where $r$ is the distance from the stellar centre, $w$ is the distance from the vertical axis (a cylindric radius), $v_*$ is the velocity at the stellar surface, $v_{esc}$ is the escape velocity, $U_*$ is the rotational velocity at the stellar equator, and $R_*$ is the stellar radius. The electron temperature $T_e$ was chosen following the law

\begin{equation}\label{tm}
    T_e(r) = T_e(R_*) (r/R_*)^{-1/\alpha}
,\end{equation}
where $\alpha$ is a parameter and $T_e(R_*)$ the electron temperature at the stellar surface.

Since the gas temperature of accreting matter near the stellar surface can also be large because of heating by stellar radiation and ultraviolet and X-ray radiation from hotspots (Dodin \cite{dod18}). Therefore, we also took into account broadening of line profiles due to the Stark effect in the MA emitting region.

\subsection{Radiative transfer}
\label{subsec:transfer}

For all emitting volumes around the star we performed non-LTE modelling of the radiative transfer. The algorithm of the calculations is described in detail by Grinin \& Tambovtseva (2011). The solution of the radiative transfer problem was based on the numerical code developed for media with large velocity gradients. The source functions were calculated in the Sobolev approximation (Sobolev \cite{sob60}), and the intensity of the
radiation emergent at frequencies within a spectral line was calculated by exact integration over spatial coordinates in the approximation of full redistribution over frequency in the co-moving coordinate system. We considered 15 atomic levels plus continuum. An inclination angle $i$ is measured from the symmetry axis  (for $i = 0$ the star is seen pole-on).

\subsection{Model calculations -- results}
\label{subsec:transfer}

In our calculations, we considered radiation emerging in the three first lines of the Balmer series, H$\alpha$, H$\beta$, and H$\gamma$. Our aim was not to} reproduce exact values of the maximum intensities in both peaks of the profiles because of their strong variability. For a satisfactory agreement with the observed profiles we required a match in line widths, shapes, and asymmetries, in addition to some specific observed features. For the comparison, we normalized both calculated and observed line profiles by dividing the intensities at each radial velocity by the maximum line intensity. Table~\ref{tab:range} contains ranges of the main parameters of the disk wind and the magnetospheric accretion models. 

\begin{table}
\centering
 \caption{Ranges of model parameters.}
\begin{tabular}{|c|c|}
\hline
  Parameter & Range  \\
  \hline
   Disk wind &   \\
\hline
  $\theta_1$ & 30-45$^\circ$ \\
\hline
  $w_1$ & 5 $R_*$ \\
\hline
  $w_N$ & 15 - 20 $R_*$ \\
\hline
  $\dot{M}_w$ & $(0.5 - 10) \times 10^{-8}M_\odot$ yr$^{-1}$ \\
\hline
  $\gamma$ & 2 - 3 \\
\hline
  $\beta$ & 1 - 5 \\
\hline
  Magnetospheric accretion &   \\
\hline
   $R_{out}$ & 3 - 5$R_*$ \\
\hline
   $\theta_{in}$ & 10 - 40$^\circ$ \\
\hline
   $\theta_{out}$ & 45 - 80$^\circ$ \\
\hline
   $\dot{M}_{acc}$ & $(0.5 - 5) \times 10^{-7}M_\odot$ yr$^{-1}$ \\
\hline
\end{tabular}
\label{tab:range}
\end{table}

\begin{table}
\centering
 \caption{Parameters of the disk wind model.}
\begin{tabular}{|c|c|c|c|c|c|c|c|}
\hline
 Model & $w_1 - w_N$ & $\theta_1$ & $f$ & $\gamma$ & $\dot{M_w}$ & $\beta$ & T  \\
  \hline
  &$R_*$ & degree &  &  & $M_\odot$ yr$^{-1}$ &  & K  \\
  \hline
  DW1 & 5 - 20  & 45  &  5 & 2 & $5 \times 10^{-9}$  & 5 & 8000 \\
\hline
  DW2 & 5 - 20  & 45  &  5 & 2 & $5 \times 10^{-8}$  & 5 & 8000 \\
\hline
\end{tabular}
\label{tab:wind}
\end{table}

\begin{table}
\centering
\caption{Parameters of the magnetospheric accretion model.}
\begin{tabular}{|c|c|c|c|c|c|}
\hline
 Model & $R_{out}$  & $\theta_{in} - \theta_{out}$ & $T_{R_*}$ & $\alpha$ & $\dot{M}_{acc}$\\
       &  $R_*$ &  degrees &   K  & & $M_\odot$ yr$^{-1}$ \\
\hline
 MA1  &  5     &  40 - 60  & 10000 &  3   &  $5\times 10^{-7}$  \\
\hline
 MA2  &  5     &  20 - 45  & 8000  &  3  &  $5\times 10^{-7}$  \\
\hline
\end{tabular}
\label{tab:magnet}
\end{table}

As a demonstration of the details of our calculations we present in Fig.~\ref{fig:Halpha} all separate components forming the H$\alpha$ line profiles together with the resulting line profiles, one for a small inclination, 30$\degr$ and another for 65$\degr$. The radiation from the MA region is shown in red (model MA1), and the radiation from the disk wind is shown in blue (model DW1). The total line profiles are depicted in black. 

One can see that at small inclinations the disk wind provides an additional emission in spite of the fact that the mass accretion rate is large in comparison with the mass-loss rate. At large inclinations the disk wind absorbs the part of the radiation formed in MA and in the disk wind itself. This is why the lines are double-peaked in this two-component model.

Examples of how the Balmer line profiles depend on inclination angle are presented in Fig.~\ref{fig:balmer} in the Appendix. The left panel shows H$\alpha$, H$\beta$, and H$\gamma$ profiles assuming magnetospheric accretion only (model MA1). Since the rotational velocities in both components of S CrA appear to be small (see Table~\ref{parameters}) we neglect the tangential velocity component in our calculations. The right panel shows the same lines calculated with the hybrid model (model MA1 plus a disk wind model DW1). It is seen that the disk wind changes the line profiles substantially: at small inclination angles, it gives an emission at low radial velocities, {whilst at larger angles ($>$  50$\degr$) it absorbs the radiation from the MA region, and also its own radiation, giving rise to a central dip in the line profiles}. Nevertheless, the accretion features are still prominent, more clearly seen in the weaker lines (H$\beta$ and H$\gamma$).

\begin{figure}
\centering
\includegraphics[width=8.5cm, angle=-00]{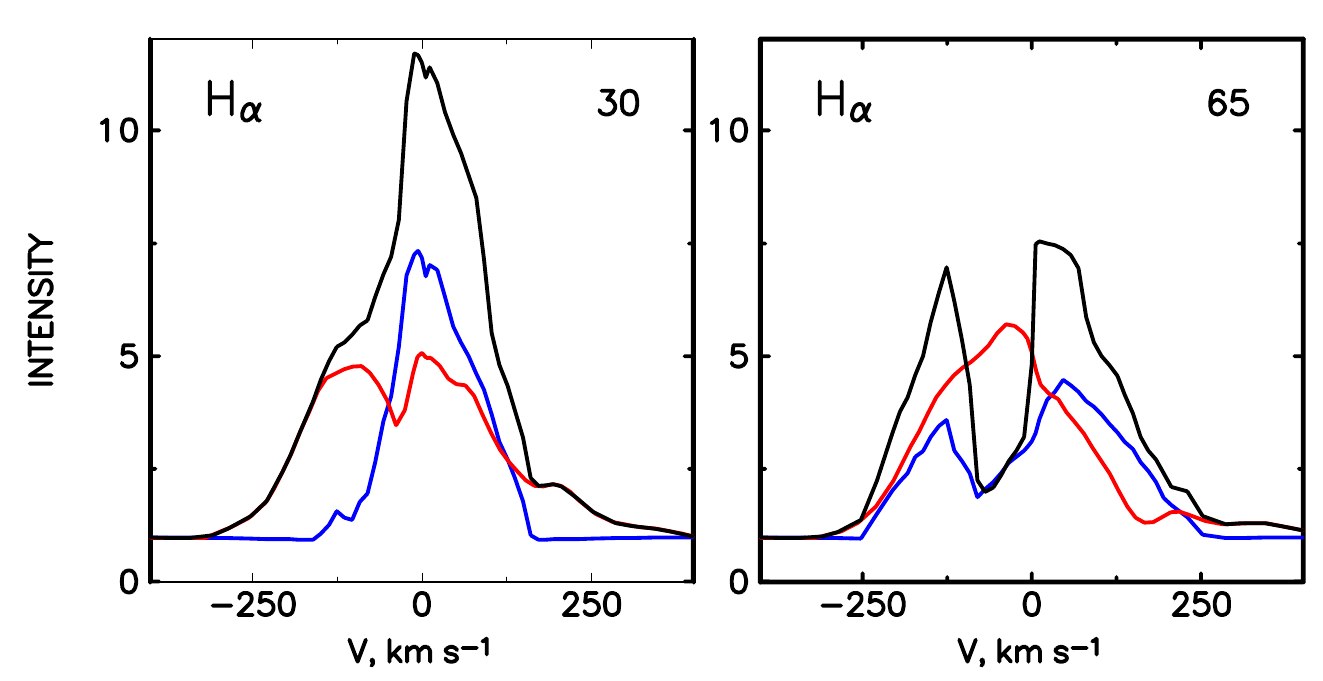}
\caption{Components of the H$\alpha$ line profiles calculated in model MA1 + DW1 at 30$^\circ$ (left) and 65$^\circ$ (right): magnetospheric accretion (red), disk wind (blue), and the resulting line profile (black).}
\label{fig:Halpha}
\end{figure}

\begin{figure}
\centering
\includegraphics[width=8.5cm, angle=-00]{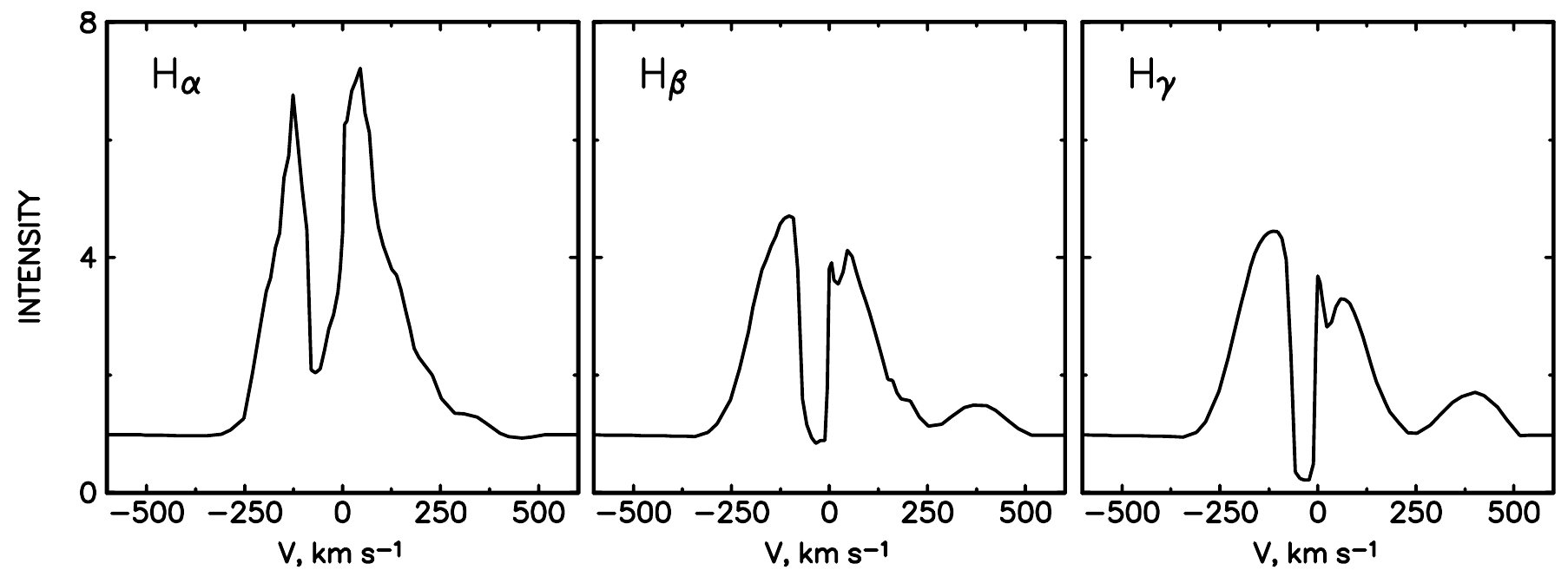}
\includegraphics[width=8.5cm, angle=-00]{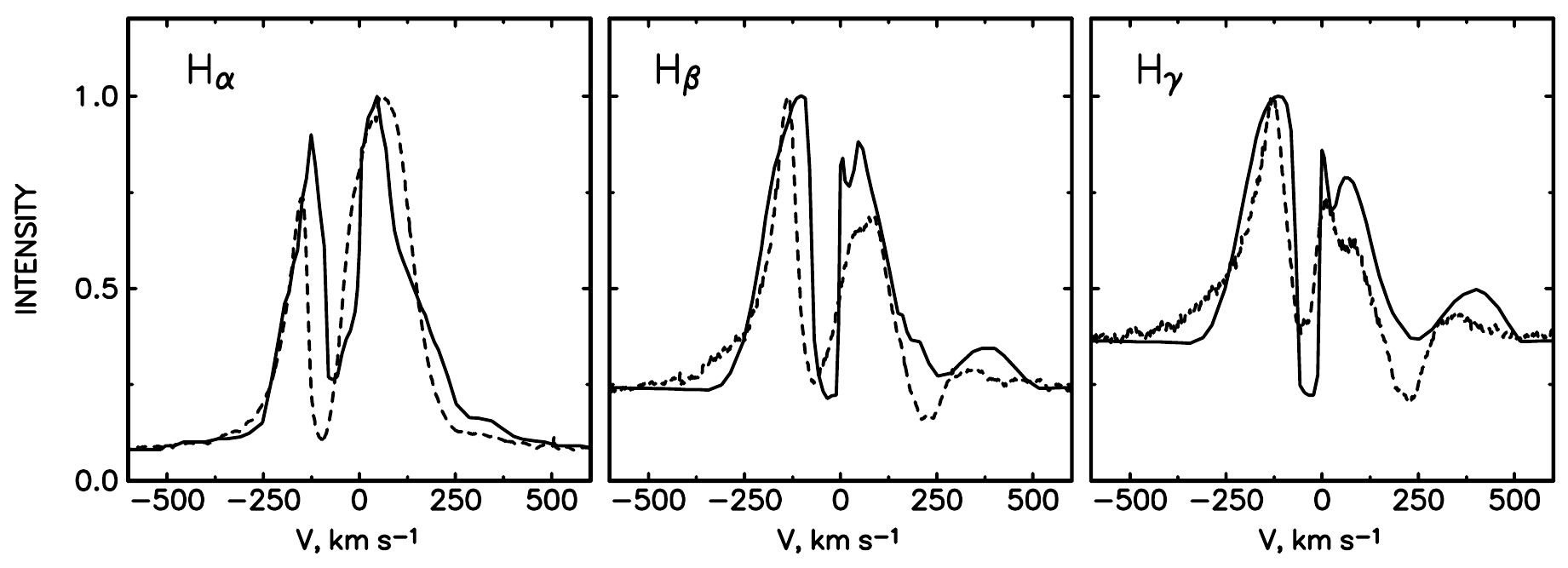}
\caption{Balmer line profiles in S CrA SE. Upper panels: H$\alpha$, H$\beta$ and H$\gamma$ line profiles as calculated from the hybrid model MA1 plus DW1 and with an inclination angle of 65$\degr$. Lower panels: The same profiles normalized to peak intensity (solid curves) superimposed on the observed profiles (dashed curves).}
\label{fig:profSE}
\end{figure}

\begin{figure}
\centering
\includegraphics[width=8.5cm, angle=-00]{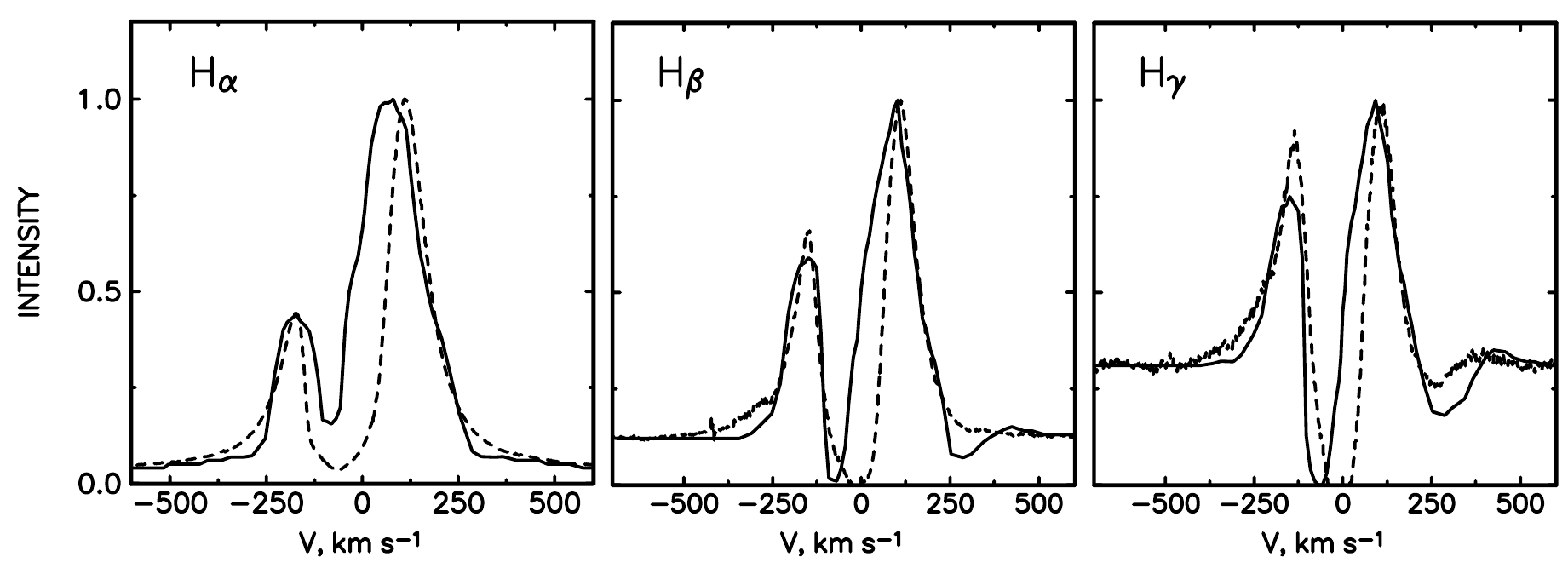}
\caption{As in Fig. 3 but  S CrA NW. The line profiles are calculated in the hybrid model MA2 plus DW2 and with an inclination angle of 
60$\degr$.}
\label{fig.profNW}
\end{figure}

Our calculations showed that the best qualitative agreement between observed and calculated line profiles for S CrA SE is reached with the combined models MA1 and DW1. Figure~\ref{fig:profSE} presents the theoretical line profiles of H$\alpha$, H$\beta$ and H$\gamma$ normalized to the continuum (top panel), and below, these are overlapped on the observed profiles. Best fits were achieved for inclination angles of $\sim$ 65$\degr$. All observed features are accounted for in the model, for instance the ratios between the blue-shifted and red-shifted peaks, which are similar to those observed.

The Balmer line profiles for S CrA NW differ from those in S CrA SE as can be seen in Fig.~\ref{fig.profNW}. Firstly, the wind features are revealed more clearly: the blue peak is smaller than the red one in all lines, which is typical for an outflow with rotation. Secondly, the lines are extremely strong, reaching a factor of $\sim 40$ relative to the continuum level. One explanation is that the mass-loss rate is exceptionally large. Another possibility is that the stellar continuum has been depressed due to foreground extinction from dust. In this case, the strengths (equivalent widths) of emission lines appear to increase due to contrast. Finally, the central absorption components are very strong and broad. In any case, the presence of a strong disk wind is obvious. We therefore increased the value of the mass-loss rate by an order of magnitude (model DW2) and checked different MA models. Results of these calculations are shown in Fig.~\ref{fig.profNW}, where a qualitative agreement between observed and theoretical profiles has been reached. 

Finally, we note that models of a polar wind seen nearly pole-on would produce a wide central absorption and double-peaked line profiles. However, in the simulations by Matt \& Pudritz (\cite{mat07}) and Cranmer (\cite{cra08}), where the wind is driven by the thermal pressure, there is only a small region close to the star where the gas temperature is suitable for hydrogen line formation. In addition, this wind is optically thin, and the mass-loss rate is small, $\sim 10^{-11} - 10^{-9}$ M$_\odot$ yr$^{-1}$. Our calculations showed that the line profiles are symmetric around zero velocity, and the peak intensities are small and practically equal. Thus, this emitting region cannot be considered as an important contributor to the total line radiation.

In conclusion, our calculated line profiles based on the hybrid model reproduce all main features of the Balmer line emission in both S CrA components satisfactorily.

\section{Discussion}
\label{sec:discussion}

Earlier photometric and spectroscopic studies of S CrA were related to the unresolved binary. A stellar luminosity of 1.5 L$_\odot$ was derived by Hamann \& Persson (\cite{ham92}) and of $L_*$ = 1.3 L$_\odot$ by Prato et al.~(\cite{pra03}). Carmona et al.~(\cite{car07}) resolved the NE and SE components and  assigned spectral types G6 and K5, respectively. Vural et al.~(\cite{vur12}) found the following parameters for NW and SE: spectral types K3 and M0, stellar luminosities 2.3 $\pm$ 0.7 and 0.76 $\pm$ 0.24 L$_\odot$. This contradicts our results indicating that the two stars are very similar. It is likely that the earlier spectral type estimated for NW as compared to SE is a consequence of the highly veiled spectrum of NW. Our high-quality and high-resolution spectra enable more accurate determination of the spectral types.

The Balmer line profiles are different in NW and SE, and this fact may help us to understand which component was dominant in the spectra taken of the unresolved star. Although NW is usually referred to as the primary component, the high-resolution spectrum of S CrA in Hamann \& Persson~(\cite{ham92})  is actually that of the SE component. The H$\alpha$ line profile reported by Reipurth et al.~(\cite{rei96}) in their atlas of pre-main sequence stars also corresponds to the SE component.

From our theoretical modelling of the two S CrA stars in Section~\ref{sec:models} we could find a good match between calculated and observed Balmer line profiles when considering a combination of a magnetospheric accretion flows and a disk wind, that is, the hybrid model. Best fits were obtained for inclination angles of $\sim$ 65$\degr$ for SE and $\sim$ 60$\degr$ for NW (for $i = 0$ the star is seen pole-on). Hence, not only are the two stars very similar in radius, luminosity, and surface temperature (Section~\ref{sec:photospheric}), but the calculations indicate that the disks also have similar tilts relative the observer.  

In this case, one expects from Fig.~\ref{fig:sketch} that the lines-of-sight to both stars pass the trajectories of accreting gas falling at velocities close to the free-fall velocity just above the stellar surface. This is exactly what we observe in SE, where the red-shifted absorption components extend to velocities of + 380 km s$^{-1}$, which is the free-fall velocity at the stellar surface given the stellar parameters. We note that the red-shifted edge of the absorption line has the same velocity in all our spectra taken over the period studied. Furthermore, with the $v$ sin $i$:s presented in Table~\ref{parameters} and inclinations of $\sim$ 60$\degr$, the equatorial rotational velocities of the stars would be $\sim$ 13 km s$^{-1}$. With stellar radii of $R_*$ = 1.4~R$_{\odot}$ the periods of rotation are 5.4 days, which is typical for young TTSs. This is also congruent with the result presented in Petrov et al (\cite{pet11}), where it was demonstrated that narrow chromospheric emission lines and photospheric lines show anti-phase variations in velocity caused by an offset between the emitting region and the pole of rotation. From the observed fluctuations, one would infer a rotational period of 5 to 6 days, although the observations cover only part of the cycle.

An additional argument in favour of high inclinations comes from the periodic radial velocity variations observed for the narrow \ion{He}{i} emission lines at 4471, 5875, and 6678 \AA. This emission originates in hot regions which are off-set from the pole. In Fig.~\ref{fig.HePeriod}, we have depicted the most probable periods with sinusoids for each star. The resulting periods, 4.2 $\pm$ 1 days for NW and P = 5.5 $\pm$ 1 days for SE, are congruent with the results obtained above. However, the observations span only part of the periods in question, and further observations are warranted.  In the case of low inclinations, for example, $i$ = 30$\degr$ deg, the expected period of rotation would be less than 3 days.

\begin{figure}
\centering
\includegraphics[width=7.5cm, angle=-00]{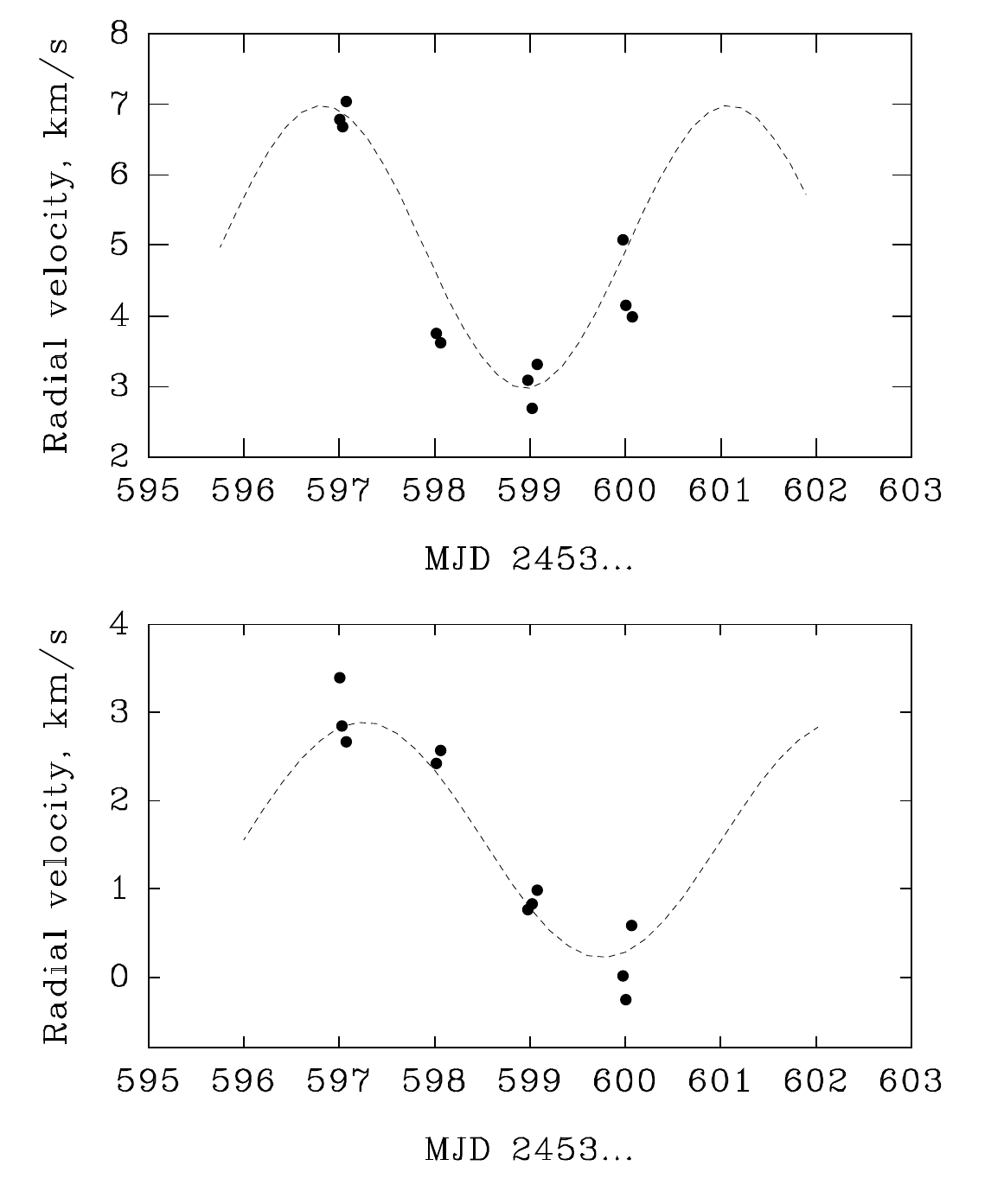}
\caption{Time variations of radial velocity of HeI emission lines. Upper panel: S CrA NW, lower panel: S CrA SE. Dashed lines: approximation by sinusoid with periods of 4.2 days (NW) and 5.5 days (SE).}
\label{fig.HePeriod}
\end{figure}
  
There are, however, some puzzling circumstances related to this scenario.
Firstly, our models suggest that the inclinations of the stars are $\sim$ 65$\degr$, and that therefore the disks are inclined to the line-of-sight by $\sim$ 25$\degr$. It follows that the occulting circumstellar grains must be located either very close to the star or high above the disk plane if at larger distances. In the first case, one may consider that dust fragments are stirred up at the innermost part of the disk, something that appears to occur in RW Aur A according to Petrov et al. (\cite{pet15}). This star undergoes large dips in brightness due to extinction changes and it has very strong red-shifted absorption components like those observed in SE. However, in RW Aur A the drops in brightness are "grey", that is, the dust grains are large, as expected from their origin in the disk plane. Another possibility, in the case of SE and NW, is that dust is carried from the inner disk edge along the accreting gas, and that dust grains sputter and become smaller on their way. 

2. From the sub-au resolution NIR maps of SE and NW, Garcia Lopez et al. ({\cite{gar17}) concluded that  the disks of both stars are seen nearly face on; the corresponding inclinations were estimated as 22$\degr$ $\pm$ 6$\degr$. and 28$\degr$ $\pm$ 3$\degr$. In line with this, Pontoppidan (2011), Schegerer et al. (\cite{sch09}), and Vural et al. (\cite{vur12}) also reported relatively low inclination angles. At these inclinations we cannot find any reasonable fit between modelled and observed line profiles. The estimates are subject to uncertainties. In our calculations, a change of $\pm$ 10$\degr$ would still give reasonable fits to the observed profiles. Furthermore, inclination angles defined from modelling of hydrogen lines and molecular CO lines can differ (e.g. Grady et al. \cite{gra10}). Estimates of disk inclinations based on different methods and extracted from different wavelength regions can be substantially different (see e.g. Eisner et al. \cite{eis04} and Doucet et al. \cite{dou06}).
  
The reason for this discrepancy remains unclear. Perturbed disks developing open spiral patterns may take forms that differ from the circular case, but it seems unlikely that the disks in both the SE and NW components would be affected in the same way. Interferometric NIR observations are sensitive to the thermal radiation from a disk wind, which according to Bans \& K\"onigl (\cite{ban12}) can give a large contribution to the NIR excess of TTSs. This additional source, if present, is not as flat as the disk and would result in lower limits of the disk inclinations in both components. Observations using the sub-millimetre interferometer ALMA can shed more light on the precise forms of the disks.         

Alternatively, the concept of standard magnetospheric accretion depicted in Fig.~\ref{fig:sketch} must be revised. To match both the low inclinations and observed accretion signatures, one might consider more complex configurations where the magnetosphere is tilted relative to the disk plane or multipole magnetic structures.

In the case where the disks are oriented nearly face-on, there is a problem in accounting for the results obtained from the rotational periods and the variable circumstellar extinction. In this case, the brightness changes might be caused by dust carried along with the disk wind or caused by dusty cloudlets containing small grains passing the stars, possibly in remote remnants of the parent cloud surrounding the binary, explanations one might consider also in our configuration. 

Nevertheless, from the present study we find that with inclinations of $\sim$ 65$\degr$ we not only get a satisfactory fit between modelled and observed Balmer line profiles, but also an agreement between the resulting rotational periods and those inferred from the anti-phase changes described above. In addition, we have found an explanation of the extent of the red-shifted absorption components in SE. These features are weak in NW, while the strength of the H$\alpha$ emission indicates a very high accretion rate. The absence of inverse P Cygni absorptions does not imply a low accretion rate however, since their strength also depends on filling factors, the inner radius of the accretion disk, and the orientation and structure of the magnetosphere.

\section{Conclusions}
\label{sec:conclusions}

We have investigated the properties of the two stars forming the visual T Tauri binary S CrA based on the results from spectral and photometric monitoring and theoretical modelling of their magnetospheric structures and winds. Both stars show large brightness variations ($\geq$ 2 mag in V band), an unusually high and variable degree of veiling, remarkable emission line spectra with both broad and narrow emission lines, and broad red-shifted absorption lines flanking strong lines, properties that are rare among the T Tauri stars. The stars are labelled NW and SE, and this study leads to the following conclusions:

We have derived a number of stellar properties and found that the two stars are remarkably similar in all relevant parameters: $T_{eff}$ (K) = 4250 K, log $g$ = 4.0, $v$ sin $i$ = 12 km~s$^{-1}$, $M_*$ = 1~M$_{\odot}$, $L_*$ = 0.7~L$_{\odot}$, and $R_*$ = 1.4~R$_{\odot}$, valid for both stars. Hence, S CrA NW and SW represents a T Tauri twin. However, their emission spectra are different.
Both stars have narrow emission lines superimposed on broad emission lines, and the latter are strongest in NW. This star also has the highest degree of veiling reaching close to VF = 10 on occasion, in terms of the standard veiling factor. As demonstrated in previous studies the contribution from narrow emission lines filling in the photospheric lines dominates in VF, especially when the veiling is high. Therefore, variations in the continuous excess emission contribute only to a smaller extent in the light changes.
The narrow emission lines are formed in surface areas off-set from the pole and vary in radial velocity in anti-phase with the photospheric lines. From these fluctuations we infer rotational periods of $\approx$ 4.2 days for NW and $\approx$ 5.5 days for SE, and combined with the measured $v$ sin $i$:s we obtain inclinations of the stars of around 60$\degr$.

Both stars have strong emission lines of neutral and ionized elements with absorption components related to both accretion flows and winds. The red-shifted absorption components are strongest in SE, extending to + 380 km~s$^{-1}$, the estimated free-fall velocity at the surface.    

In our model simulations of the gas emission in NW and SE we took into account magnetospheric accretion, a polar wind, and a disk wind. We compared the resulting Balmer line profiles with those observed and found that very good agreements can be obtained for inclinations of $\sim$ 65$\degr$ in a hybrid model, including magnetospheric accretion and a disk wind only. From our models we derive mass-loss rates in winds ranging from 5 10$^{-9}$ to 5 10$^{-8}$ $M_\odot$ yr$^{-1}$ and mass accretion rates of $\sim$ 5 10$^{-7}$. The accretion rates appear to be larger in NW by a factor of approximately two judging from spectral signatures.

For inclinations of $\sim$ 65$\degr$ the disks are inclined by $\sim$ 25$\degr$ to the line-of-sight. At this orientation the trajectories of the accreting funnels just above the stellar surface are parallel to the line-of-sight in agreement with our observation of gas moving towards SE at free-fall velocity. Moreover, the inclinations inferred from the anti-phase velocity variations mentioned above agree with those adopted in the model.     

The brightness changes are caused mainly by rapid variations in the amount of dust in the line-of-sight. The stars become redder with decreasing brightness, indicating that the foreground occulting dust grains are small, and it is unlikely that this dust has been stirred up from the inner disk. We discuss whether sputtering of larger grains carried along the accretion funnels can produce the observed fluctuations. 

Another open issue is that the inclinations derived from our observations and model calculations are different from those obtained from interferometric NIR spectroscopy, with disks oriented almost face-on, and we discuss possible causes for this puzzling discrepancy. In the case where the disks are seen nearly face-on, it is more likely that the occulting dust is carried by the disk wind or resides in dusty cloudlets further out from the system. Future observations with ALMA can shed more light on the shapes and orientations of the disks in question.

\begin{acknowledgements}

This work was supported by the INTAS grant 03-51-6311, the Swedish National Space Board, and the Magnus Bergvall foundation. FMW thanks past Provosts and Vice Presidents for Research of Stony Brook University for providing support for the SMARTS partnership.

\end{acknowledgements}


\appendix
\label{AppendixA}
\section{Hydrogen line profiles in models assuming magnetospheric accretion and a disk wind}

\begin{figure*}
\centering
\includegraphics[width=9cm]{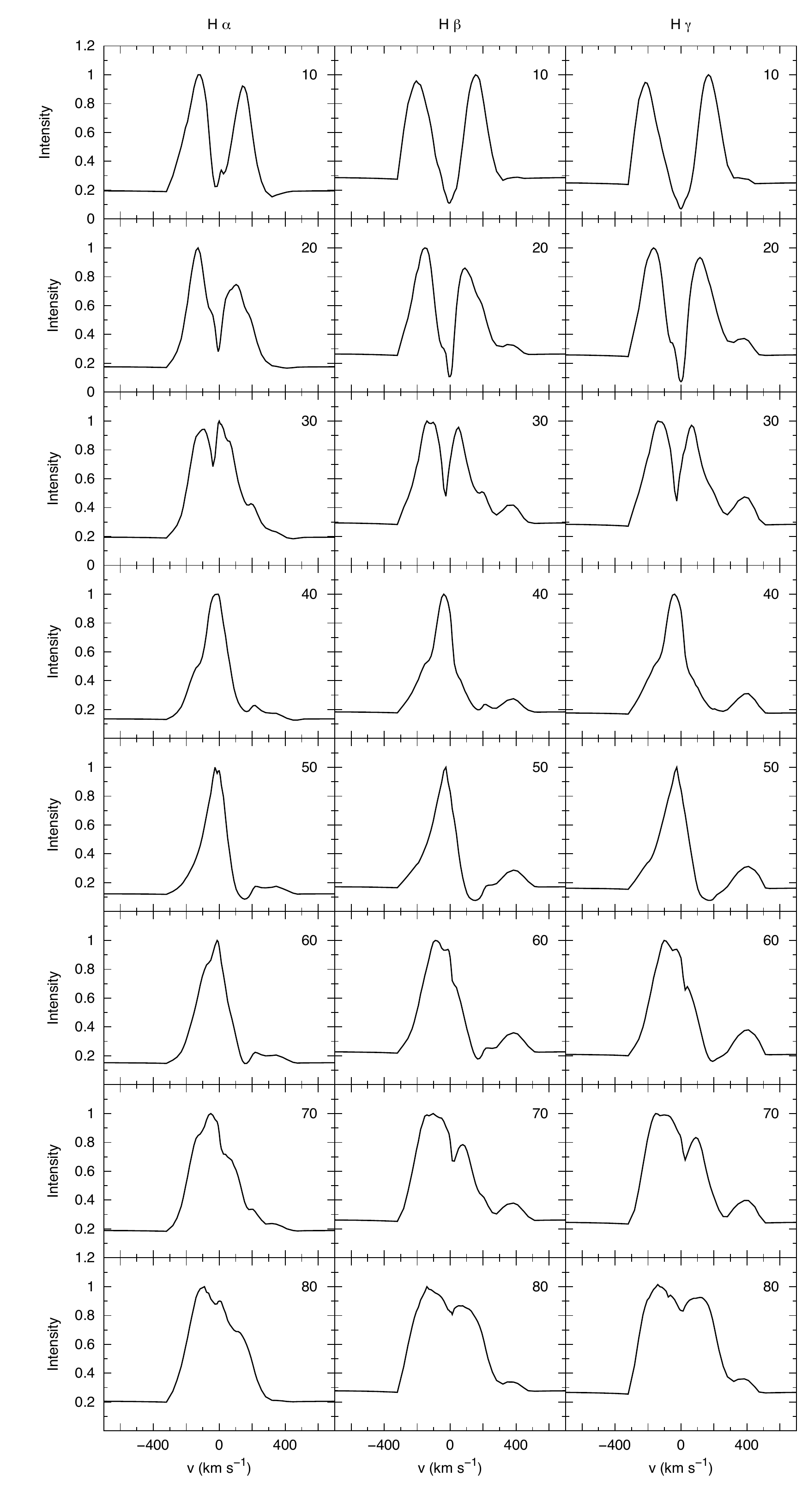}
\includegraphics[width=9cm]{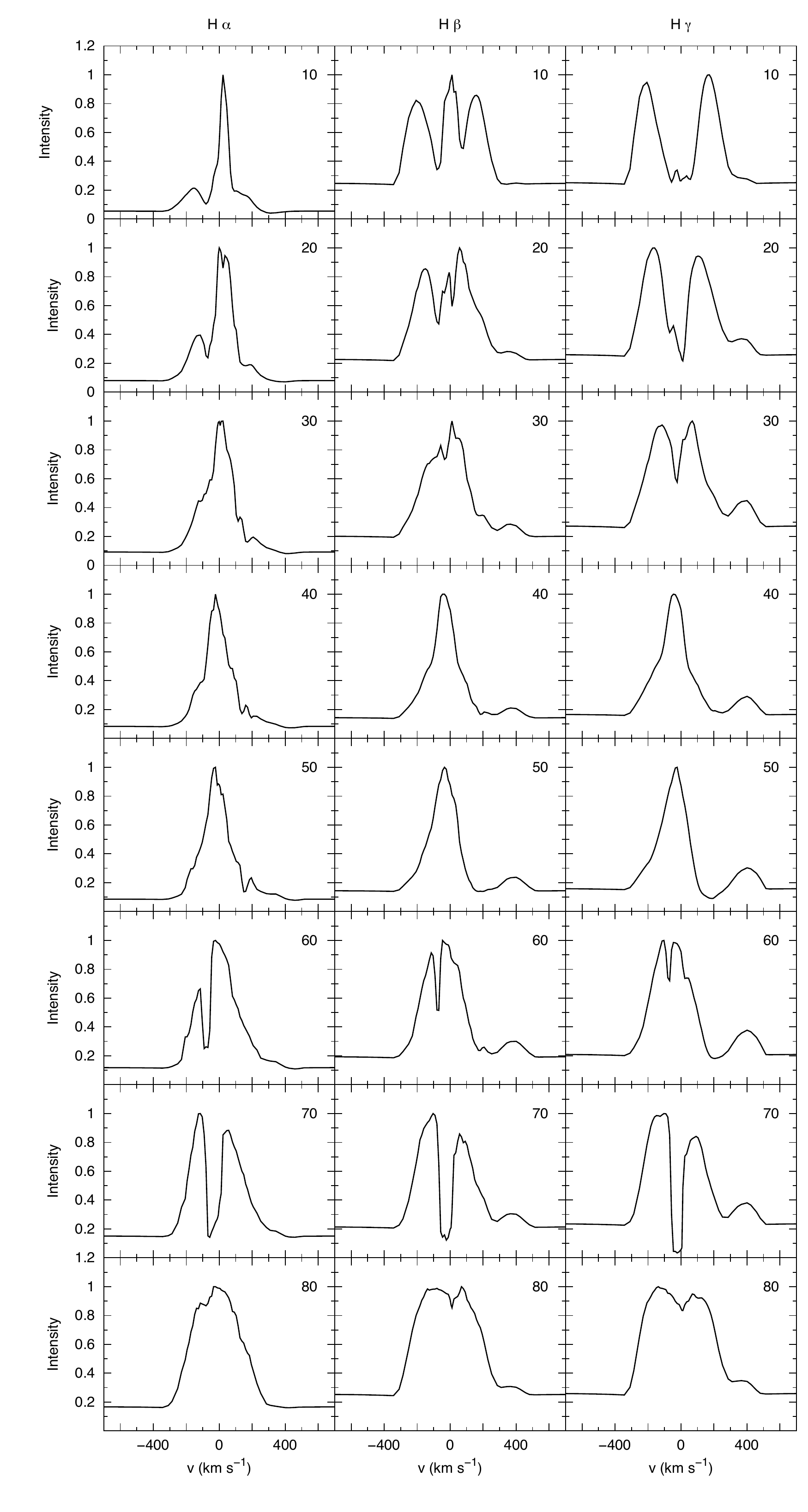} 
\caption{\textit{Left:} H$\alpha$, H$\beta$ and H$\gamma$ line profiles resulting from the magnetospheric accretion region (model MA1); \textit{Right:} As in the left panel but in the framework of the hybrid model combining magnetospheric accretion (MA1) and a disk wind (DW1). Inclination angles are marked.}
\label{fig:balmer}
\end{figure*}

Figure~\ref{fig:balmer} shows results of calculations of the Balmer line profiles in models of the magnetospheric accretion and in the hybrid model combining the magnetospheric accretion and a disk wind (models MA1 and DW1). The line profiles are shown for inclinations from 10 to 80 degrees.

\end{document}